\documentclass[aps,prb,twocolumn,epsfig,floats]{revtex4}

\usepackage{graphicx,epsf}

\usepackage{bm}       
\usepackage[parfill]{parskip}    		

\usepackage[title]{appendix}

\def\be{\begin{equation}}
\def\ee{\end{equation}}
\def\bea{\begin{eqnarray}}
\def\eea{\end{eqnarray}}

\def\avg#1{\left< #1 \right>}

\newcommand{\figwidth}{ \columnwidth}
\newcommand{\tabfigwidth}{ 1.04 \columnwidth}

\newcommand{\vect}[1]{\boldsymbol{\mathbf{#1}}}

\def\Tr{\text{Tr} \,}

\def\of#1{\left( #1 \right)}
\def\set#1{\left\{ #1 \right\}}
\def\dag{^{\dagger}}

\def\state#1{\left| #1 \right>}
\def\cstate#1{\left< #1 \right|}

\begin{document}

\title{Kondo screening in p-type two dimensional transition metal dichalcogenides}
\author{Michael Phillips and Vivek Aji}
\address{Department of Physics \& Astronomy, University of California,
Riverside, CA 92521}
\begin{abstract}
Systems with strong spin orbit coupling support a number of new phases of matter and novel phenomena. This work focuses on the interplay of spin orbit coupling and interactions in yielding correlated phenomena in two dimensional transition metal dichalcogenides. In particular we explore the physics of Kondo screening resulting from the lack of centro-symmetry, large spin splitting and spin valley locking in hole doped systems. The key ingredients are i) valley dependent spin-momentum locking perpendicular to the two dimensional crystal; ii) single nondegenerate Fermi surface per valley, and iii) nontrivial Berry curvature associated with the low energy bands. The resulting Kondo resonance has a finite triplet component and nontrivial momentum space structure which facilitates new approaches to both probe and manipulate the correlated state. Using a variational wave function and the numerical renormalization group approaches we study the nature of the Kondo resonance both in the absence and presence of circularly polarized light. The latter induces an imbalance in the population of the two valleys leading to novel magnetic phenomena in the correlated state.
\end{abstract}
\maketitle

\section{Introduction}

Single layer Transition metal Group-VI Dichalcogenides (TMDs), MX$_{2}$ (M = Mo, W; X=S, Se, Te), are direct band gap semiconductors whose physics is strongly influenced by spin orbit coupling. While they share the hexagonal crystal structure of graphene, they differ in three important aspects: 1) The spectrum possesses gaps at the $K$-points as opposed to Dirac nodes; 2) Broken inversion symmetry and coupling of the spin with momentum result in a large splitting of the valence bands; and 3) The two bands near the chemical potential arise from the partial filled transition metal d-orbitals\cite{mak1,mak2,xiaodi,zcs}. A striking consequence is the nontrivial Berry's phase of the low energy bands. The symmetry of the system is such that the $z$ component of spin, $s_{z}$ (i.e. component perpendicular to the MX$_{2}$ plane) is conserved. Associated with each band is a Berry curvature, $\vect{\Omega}$, whose $z$ component changes sign going from one valley to the other, and also when going from the conduction to the valence band. These properties allow for coupled valley and spin phenomena\cite{xiaodi,niurmp}.

Of particular significance is the ability to manipulate the valley degree of freedom. The Berry curvature engenders an intrinsic angular momentum associated with Bloch wave functions \cite{changniu}, which in turn allows for spin preserving  transitions between valence and conduction bands induced by optical fields even though the atomic orbitals involved all have $d$ character. Furthermore, the valley dependent sign of the Berry curvature leads to selective optical excitation where right circular polarization couples to one valley and left to the other. As a consequence a number of valleytronic and spintronic applications are enabled and have attracted a lot of attention over the last few years \cite{xyn,xuwse2,niurmp,xiaodi,makvh}.

\begin{figure}
\includegraphics[width=\linewidth]{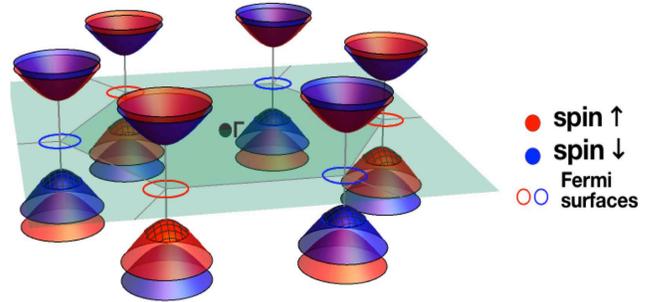}
\caption{Low energy band structure of hole doped TMD. The unique aspect of the system is the spin split band allowing for spin nondegenrate Fermi surfaces around the K points. }
 \label{mosbs}
\end{figure}

While much of the focus thus far has been on the nontrivial properties engendered in the noninteracting limit, our work emphasizes the band structure and valley contrasting probe to study and manipulate correlated phenomena in these systems. This is particularly interesting in hole doped systems where an experimentally accessible window in energy is characterized by two disconnected pieces of spin nondegenerate Fermi surfaces (see fig.\ref{mosbs}). (Note that these considerations do not apply to MoS$_{2}$ because there is also a spin degenerate Fermi surface at the $\Gamma$ point.) Since one can preferentially excite electrons from one or the other Fermi surface, optical probes have spin specificity. These features have important implications on magnetic phenomena. Here we focus on the nature of the Kondo effect employing two methods namely the variational wave function\cite{yosidakondo,yafetvarma}, and the numerical renormalization group (NRG)\cite{wilsonnrg,kmww1,bcpnrg}. 

In conventional metals the screening of an isolated magnetic impurity relies on the formation of a singlet state between the impurity moment and the electrons. In TMDs the up and down spins at the same energy occupy different valleys and spin flip scattering requires inter valley processes. For a hole doped system, where the chemical potential intersects only one of the two spin split bands, the main findings are as follows: i) the Kondo resonance is an equal admixture of singlet and the $m=0$ triplet formed between the impurity spin and the band fermions. Thus the ground state has total spin $(\vect{S} + \vect{\sigma})^{2} = J^{2} \approx \hbar^{2}$, where $\vect{S}$ and $\vect{\sigma}$ are spin operators for the impurity and band fermions respectively. The result reflects the fact that the Kondo energy scale as well as the level broadening  are small compared to the spin splitting scale which is $\sim$ 0.1-0.5 eV; ii) the hybridization of the impurity state with the TMD depends crucially on the interplay between the symmetry of the atomic orbitals involved and band topology (i.e. Berry curvature). To illustrate, orbitals with similar atomic overlap with the fermionic states have different hybridization due to the additional orbital angular momentum associated with the Berry curvature; iii) the spin specificity of opto-electronic coupling allows access to components of the screening cloud which in turn allows for the tuning of spin in the Kondo state.  

We introduce the model for Anderson impurities in monolayer TMDs (sec. \ref{sec:model}).
The results of the spin structure are obtained using a variational wave-function approach (sec. \ref{sec:varfunc}) for a single impurity located on top of a M (Mo or W) site. The location is chosen for illustrative purposes to show the interplay of topology and interaction, and other high symmetry sites for the impurity only modify the precise form of the hybridization. Since the density of states is always finite in the hole doped systems, the energetics and stability of the Kondo resonance are rather conventional. What is striking however is the composition of the Kondo cloud and its physical properties as emphasized above.
Additionally, to understand the effect of spin/valley specific optical coupling, we find that the variational Kondo ground state starting from an optically excited Fermi sea is insufficient and a better picture is obtained from NRG (sec. \ref{sec:nrg}).
In this case we find that $J_{z}\neq0$ and scales with the strength of optical field excitation, revealing the tunability of the correlated spin state with optical probes. Prior to concluding we compare the variational and NRG methods and results (sec. \ref{sec:compare}).

\section{Model}		\label{sec:model}

In this section we describe the material and impurity model Hamiltonians.

\subsection{Low energy bands and topology}
The minimal model for TMDs in the low energy bands near $\vect{K}$ ($\vect{K}'$)  is in terms of the two basis functions 
$\left|1\right> =  \left|d_{z^{2}}\right>$ and 
$\left|2\right>= \left( \left|d_{{x^{2}-y^{2}}}\right>+i\tau   \left|d_{xy}\right>\right)/{\sqrt{2}}$, 

\begin{equation}  \label{eq:metalHamil}
H_{m} = a t\left(\tau\sigma_{x}k_{x}+\sigma_{y}k_{y}\right)+{\Delta\over 2}\sigma_{z}-\lambda\tau{\sigma_{z}-1\over 2}s_{z}
\end{equation}

where $\sigma_{i}$ are Pauli matrices in the space of two bands represented by the eigenvalues $\pm 1$ of $\sigma_{z}$, $\tau = \pm 1$ is the valley index, $s_{z}$ is the Pauli matrix for spin, $a$ is the lattice constant, $t$ is the effective hopping parameter, $\Delta$ is the gap at the $K$ points and $\lambda$ is the spin orbit coupling \cite{xiaodi,zcs}.  Written in terms of the magnitude $k=|\vect{k}|$ and azimuthal angle $\phi=\arctan{(k_y/k_x)}$, the Hamiltonian (\ref{eq:metalHamil}) is diagonalized by the unitary matrix $U(\vect{k},\tau,s)$:

\begin{equation}  \label{eq:wMat1}
U(\vect{k},\tau,s) = \left( \begin{array}{c c}  {  \chi_{k,\tau,s}} &  w_{k,\tau,s}  \\ \tau w_{k,\tau,s} { \text{e}  }^{ i\tau \phi  }  & -\tau\chi_{k,\tau,s} { \text{e}  }^{ i\tau \phi  }  \end{array} \right) 
\end{equation}

with $\chi_{k,\tau,s} = \cos{\of{\theta_{k,\tau,s}/2}}$, $w_{k,\tau,s}=\sin{\of{\theta_{k,\tau,s}/2}}$ and $ \cos{\of{\theta_{k,\tau,s}}} = { \of{{\Delta} - {\lambda \tau s }} /  \sqrt{ ({\Delta} - {\lambda \tau s })^{2} + (2 a t k)^{2}  }    }$ where $s=\pm$ for eigenvalues of $s_{z}$. The eigenvalues, the diagonal elements of $U^{\dagger} H_{m} U$ labeled $n=\pm1$, are given by $E_{n,k,\tau,s}={1 \over 2}(\lambda s \tau+n\sqrt{(2atk)^2+(\Delta-\lambda s \tau)^2})$. The Berry curvature is encoded in $\theta_{k,\tau,s}$. Mapping $(\vect{k}, E_{+,k,\tau,s})\rightarrow (\phi, \tau\theta_{k,\tau,s})$ and $(\vect{k}, E_{-,k,\tau,s})\rightarrow (\phi, \tau\theta_{k,\tau,s} - \tau\pi)$ wraps the conduction and valence bands (respectively) onto half the Bloch sphere with the texture of a skyrmion (either chiral or hedgehog).

\subsection{Hybridization}
To study the nature of Kondo screening in a hole doped system, where the chemical potential is in the topmost spin split valence band, we introduce an impurity orbital on top of the  M atom. The choice is for simplicity and does not affect the results as long as a site of high symmetry is chosen. For generality however, this section does not presume a specific location --- the on-site choice is evaluated and employed in following sections. The magnetic impurity brings its own orbital with Coulomb repulsion, $H_{imp} = \sum_{s} \varepsilon_0 f_{s}^{\dagger}f_{s} + U n_{f,\uparrow}n_{f,\downarrow}$. Next, since the Hamiltonian  (\ref{eq:metalHamil}) is only an effective low energy theory, we impose an upper cutoff in energy, $\Lambda$, within which its hybridization with the impurity takes the form \cite{andersonkondo,uchoakondo,chaokondo}

\begin{equation}  \label{eq:hyb1}
H_{V} = \sum_{\alpha,\tau,s} \sum_{j} \left(V_{\alpha, \tau, j} a_{\alpha,\tau,s}^{\dagger}(\vect{r}_{j})f_{s} + h.c.  \right)  
\end{equation}

where $a_{\alpha,\tau,s}^{\dagger}(\vect{r}_{j}) = N_{M}^{-1/2}  \sum_{\vect{k}} a_{\alpha,\vect{k},\tau,s}^{\dagger} \text{e}^{- i \vect{k}\cdot\vect{r}_{j} }$ is the creation operator at M site $\vect{r}_{j}$ with $\alpha =1 \text{ or } 2$ corresponding to the basis states. The sum over $j$ runs over the M nearest to the impurity site, $f_{s}$ is the annihilation operator of the localized electron on the impurity (taken as the origin), and $V_{\alpha, \tau, j}$ is the hybridization strength between the localized orbital with the $\left| 1 \right>$ and $\left| 2 \right>$ orbitals on the M atom at site $j$. The number of metal sites (or unit cells) is $N_{M}$.

To analyze the screening of impurity moments we must first project to the eigenspace, $a_{\alpha,\vect{k},\tau,s} = \sum_{n=\pm} U_{\alpha,n}(\vect{k},\tau,s) c_{n,\vect{k},\tau,s}$. Generally, this means that the TMD and hybridization Hamiltonians are simply

\begin{equation}  \label{eq:metalHmod2}
H_m =  \sum_{n,\vect{k},\tau,s} \thinspace E_{n,k,\tau,s} c_{n,\vect{k},\tau,s}^{\dagger} c_{n,\vect{k},\tau,s}  
\end{equation}
\begin{equation}  \label{eq:hyb3}
H_{V} = {1 \over \sqrt{N_{M}} } \sum_{n,\vect{k},\tau,s} \thinspace \left( v_{n,\vect{k},\tau,s} c_{n,\vect{k},\tau,s}^{\dagger} f_{s} + h.c. \right)  
\end{equation}

where $ v_{n,\vect{k},\tau,s} = \sum_{j} \of{\sum_{\alpha=1,2} U^{\dagger}_{n,\alpha}(\vect{k},\tau,s) V_{\alpha,\tau,j} } { \text{e}  }^{ -i \vect{k}\cdot\vect{r}_{j}  }$. Owing to the structure of the $U$ matrices from (\ref{eq:wMat1}), terms with no angular dependence pair with the $\alpha=1$ orbital whereas those with the dependence $\sim { \text{e}  }^{ -i \tau \phi  }$ pair with the $\alpha=2$ orbital.

To address the angular dependence, a discussion which is especially useful in the context of NRG \cite{kmww1}, we transform to a quasi-angular momentum (QAM) basis (termed so because $\phi$ is defined with respect to the $K$ rather than the $\Gamma$ point). The  $\vect{k}$-dependence is recast as the magnitude $k$ and a quasi-angular momentum index $\nu \in (-\infty, \infty)$ \cite{zitkokondo}. The sums become $\sum_{\vect{k}} \to {N_{M} \Omega_{c} / (2\pi)^{2} } \int d^{2}k$ and the eigenstate operators in the QAM basis are obtained with $c_{n,\vect{k},\tau,s} = ({N_{M} \Omega_{c} k / 2\pi })^{-1/2}  \sum_{\nu} \text{e}^{i \nu \phi}  c_{n,k,\nu,\tau,s}$, where $\Omega_{c} = (\sqrt{3}/2) a^{2}$ is the area of a unit cell. The Hamiltonians transform to

\begin{equation}  \label{eq:metalHmod3}
H_m =  \sum_{n,\nu,\tau,s} \int dk \thinspace E_{n,k,\tau,s} c_{n,k,\nu,\tau,s}\dag c_{n,k,\nu,\tau,s}  
\end{equation}

\be
H_{V} =  \sum_{n,\nu,\tau,s} \int dk \sqrt{ \Omega_{c} k \over 2\pi} \left(  v_{n,k,\nu,\tau,s}  c_{n,k,\nu,\tau,s}^{\dagger} f_{s} + h.c. \right)  
\label{eq:hyb4}
\ee

with an effective coupling $v_{n,k,\nu,\tau,s}= \int {(d\phi / 2 \pi)} v_{n,\vect{k},\tau,s} { \text{e}  }^{- i \nu \phi  }$ to the impurity.  This form is related to the original $V_{\alpha,\tau,j}$ appearing in eq. (\ref{eq:hyb1}) as demonstrated by expanding in Bessel functions, 

\bea
v_{n,\vect{k},\tau,s} &=& \sum_{\alpha=1,2} U^{\dagger}_{n,\alpha}(\vect{k},\tau,s) \sum_{\nu} { \text{e}  }^{ i \nu \phi  } \overline{V}_{\alpha,k,\nu,\tau}  \nonumber \\
\overline{V}_{\alpha,k,\nu,\tau} &=& \sum_{j} V_{\alpha, \tau, j} (-i)^\nu J_{\nu}(k r_{j}) { \text{e}  }^{- i \nu\phi_{j} } 
\label{eq:bessel}
\eea

where $\phi_{j}$ is the real-space angle from the position $\vect{r}_{j}$ of the M atom. 
Since the matrix is factorized as
\bea
U(\vect{k},\tau,s) &\equiv& M(\phi,\tau) \cdot N(k,\tau,s)   \nonumber		\\
&=& \left( \begin{array}{c c}  1 &  0  \\  0 & { \text{e}  }^{ i\tau \phi  } \end{array} \right) \cdot \left( \begin{array}{c c}  {  \chi_{k,\tau,s}} &  w_{k,\tau,s}  \\ \tau w_{k,\tau,s}   & -\tau\chi_{k,\tau,s}   \end{array} \right)
\label{eq:wMat3}
\eea

the hybridization  takes the form

\begin{widetext}
\begin{equation}  \label{eq:hyb5}
H_{V} =  \sum_{\nu,\tau,s} \int dk \sqrt{\Omega_{c} k \over 2\pi} 
 \left\{ \left( \begin{array}{c}  \overline{V}_{1,k,\nu,\tau} \\ \overline{V}_{2,k,\nu,\tau}  \end{array} \right)^{T} \cdot \left( \begin{array}{c} \chi_{k,\tau,s}  c_{+,k,\nu,\tau,s}^{\dagger} + w_{k,\tau,s}  c_{-,k,\nu,\tau,s}^{\dagger}   \\  \tau w_{k,\tau,s} c_{+,k,(\nu-\tau),\tau,s}^{\dagger} -\tau\chi_{k,\tau,s} c_{-,k,(\nu-\tau),\tau,s}^{\dagger} \end{array} \right) f_{s} + h.c. \right\}.
\end{equation}
\end{widetext}

Note that the interaction $V_{1, \tau , j}$ couples only to the host states with quasi-angular momentum $\nu$, while $V_{2, \tau, j}$ couples only to the states with $\nu-\tau$. Interestingly, this pattern originates from a gauge choice in the diagonalization of (\ref{eq:metalHamil}) so we expect physical quantities to be independent of the actual value of the QAM labels.

The above results reduce to the case of graphene and topological insulator when the gap and spin splitting vanish, $\Delta=\lambda=0$ \cite{zitkokondo,uchoakondo,ajikondo}. The key new aspect due to the Berry curvature is the dependence of the hybridization on the angle $\theta_{k,\tau,s}$ which encodes the nontrivial topology of the states, in addition to the orbital wave-function overlap that determines $V_{\alpha,\tau,j}$ and $\overline{V}_{\alpha,k,\nu,\tau}$.

\subsection{Impurity at M site}
We focus on the Kondo effect for magnetic adatoms, where the impurity state is on the M site and the overlap (\ref{eq:hyb1}) is with only one M atom. Symmetry mandates two classes: i) orbitals of type I defined as $s, p_{z}, d_{z^{2}}$ and $f_{5z^{3}-3zr^{2}}$ couple to orbital $\left| 1 \right>$, and ii) type II orbitals $d_{x^{2}-y^{2}}, d_{xy}, f_{zx^{2}-zy^{2}}$ and $f_{zxy}$ couple to orbital $\left| 2 \right>$. Therefore the hybridization strength $V_{\alpha, \tau, j}$ is nonzero for $\alpha=1 \text{ or } 2$ but not both, implying $H_V$ enters with trivial angular dependence. Since the adatom orbitals have maximum overlap with the nearest M site the hybridization strength is $V_{\alpha, \tau, j} = V_{\alpha, \tau} \delta_{\vect{r}_{j},0}$. Also, since we are interested in hole doped systems we project to the valence band. Then the simple form (\ref{eq:hyb3}) for type I becomes,

\begin{equation}   \label{eq:type1Hyb}
H_V = {1 \over \sqrt{N_M}} \sum_{\vect{k},\tau,s} \left( v^{1}_{-,\vect{k},\tau,s} c^{\dagger}_{-,\vect{k},\tau,s} f_{s} + h.c. \right)
\end{equation}

with $v^{1}_{-,\vect{k},\tau,s} = w_{k,\tau,s} V_{1}$. 
For the type II case, the corresponding definition will need $w \rightarrow -\tau \chi  { \text{e}  }^{ -i \tau \phi  } $. The Berry curvature plays a crucial role in determining the coupling: $\theta_{k,\tau,s}$ goes from the north pole of the Bloch sphere at $K$ ($K'$) to the equator as $k$ increases. Thus type II orbitals couple more strongly to the valence band than type I.

As above, for NRG it is useful to have the hybridization decomposed into QAM channels. Under the same conditions, the impurity lying on top of an M site gives $\overline{V}_{\alpha,k,\nu,\tau} = V_{\alpha,\tau,0} \delta_{\nu,0}$ so the alternate form (\ref{eq:hyb5}) for type I is

\be
H_V  = \sum_{\tau,s} \int dk \sqrt{\Omega_{c} k \over 2\pi} \left( w_{k,\tau,s} V_{1}  c^{\dagger}_{-,k,0,\tau,s} f_{s} + h.c. \right)  .
\label{eq:type1Hybalt}
\ee

This is clearly just eq. (\ref{eq:type1Hyb}) transformed to the new basis with apparently no change. On the other hand, for type II, we would see the same as above but with $w \to -\tau\chi$, as well as a shifted QAM label $0 \to -\tau$ on the operators. Thus the alternate forms (\ref{eq:type1Hybalt}) and (\ref{eq:hyb5}) are equivalent to the simpler (\ref{eq:type1Hyb}) and (\ref{eq:hyb3}), but with all angular dependence shifted to the operators with QAM labels. We will not employ this last form until the NRG setup where it is useful.

\section{Variational Wave function}		\label{sec:varfunc}

 We examine the ground-state properties using the variational wave-function approach. This method has ben employed in other systems to calculate the spin, suscpetibility, and energy scale of the Kondo effect. The full Hamiltonian is $H = H_0 + H_V$, where
$H_0 = H_m + H_{imp}$
describes the system with the impurity \cite{andersonkondo,hewson}, and $H_V$ is written in the form (\ref{eq:type1Hyb}).  All energies are measured relative to the chemical potential $\mu$.
For hole doped systems of interest, $-{\Delta \over 2} - \lambda < \mu < -{\Delta \over 2} + \lambda < 0$. 

For large Coulomb repulsion on the impurity level the variational state $\left| \psi \right>$ includes the ground state of the pure system $\left| \psi_0 \right>$ and states with a singly occupied impurity level \cite{yosidakondo,yafetvarma}. Since inversion is broken both singlet and triplet combinations must be included. Therefore, 

\begin{eqnarray} 
\left| \psi \right> &=& b_0 \left| \psi_0 \right> + \sum_{\ell} \left[ p_{\ell} (f^{\dagger}_{\uparrow} c_{\ell,\uparrow} + f^{\dagger}_{\downarrow} c_{\ell,\downarrow}) \right.  \nonumber  \\  \label{eq:varGnd1}
&& + \left. t_{\ell} (f^{\dagger}_{\uparrow} c_{\ell,\uparrow} - f^{\dagger}_{\downarrow} c_{\ell,\downarrow}) \right] \left| \psi_0 \right> 
\end{eqnarray}

where $b_0$ is the amplitude of the ground state in the absence of the impurity, $p_{\ell}$ is the singlet amplitude, and $t_{\ell}$ is the triplet amplitude, giving a total of three variational parameters. For brevity we use $\ell = \{ n, \vect{k}, \tau \}$. This state can be written more compactly by defining $B_{\ell,s}= p_{\ell} + s \thinspace t_{\ell}$:

\begin{equation} \label{eq:varGnd2}
\left| \psi \right> =  b_0 \left| \psi_0 \right> + \sum_{\ell ,s} B_{\ell ,s} f^{\dagger}_{s} c_{\ell ,s} \left| \psi_0 \right> .
\end{equation}

\subsection{Variational Parameters}		\label{sec:varpar}

We determine the variational parameters and ground-state energy when the impurity level sits below the chemical potential, $\varepsilon_{0}<0$. The energy is written as 
$\left< \psi \right| H \left| \psi \right> = (E_0 + \varepsilon_0 + \epsilon)\left< \psi | \psi \right>$
where $E_0 = \left< \psi_0 \right| H_m \left| \psi_0 \right>$, subject to the constraint $\left< \psi | \psi \right> = 1$.  The energy shift $\epsilon$ is determined by minimization, which then yields the variational parameters,

\begin{eqnarray} \label{eq:varEqs1}  
b_{0}= {1 \over \sqrt{N_{M}} } {\sum_{\ell,s}}'  { v_{\ell,s} B_{\ell,s}  \over \varepsilon_{0} + \epsilon } , ~~B_{\ell ,s} = {1 \over \sqrt{N_{M}} } { v^{\star}_{\ell,s} b_0  \over  \varepsilon_{\ell,s}  +  \epsilon } .
\end{eqnarray}

The notation $\sum_{\ell,s}' $ indicates summation over occupied states states: $E_{\ell,s} - \mu \equiv \varepsilon_{\ell,s}<0$.  Solving for $\epsilon$, 

\begin{equation}   \label{eq:varSolnE}
\epsilon = - \varepsilon_0 + {1 \over N_{M} } {\sum_{\ell,s}}' {|v_{\ell,s} |^2  \over \varepsilon_{\ell,s}  +  \epsilon}.
\end{equation}

Imposing the normalization $\left< \psi | \psi \right> = 1$, the singlet/triplet parameters are found in terms of $\epsilon$ and $b_{0}$ given by

\begin{equation}   
b_0 =  \left[ 1 + {1 \over N_{M} }{\sum_{\ell,s}}' {  |v_{\ell,s} |^2 \over  (\varepsilon_{\ell,s}  +  \epsilon)^2 }  \right]^{-1 / 2}.
\label{eq:varSolnPars}
\end{equation}

Using Eqs. (\ref{eq:varEqs1}), (\ref{eq:varSolnE}), and (\ref{eq:varSolnPars}) we state the solution in terms of the original singlet/triplet parameters: $p_{\ell }=(B_{\ell ,\uparrow}+B_{\ell ,\downarrow})/2$, and $t_{\ell }=(B_{\ell ,\uparrow}-B_{\ell ,\downarrow})/2$. In the absence of spin orbit coupling, every point in $\vect{k}$-space is doubly degenerate and we expect the singlet parameters to survive while the triplet parameters to go to zero. Indeed, for weak spin orbit coupling (i.e. $\lambda \ll \Delta$), to leading order $p_{\ell }\propto$ const. and $t_{\ell }\propto \lambda$. 

With cutoff $E_{-1,k_{\Lambda},s,s} = - \Lambda$,
 Eq.(\ref{eq:varSolnE}) to leading order is

\begin{equation}   \label{eq:varSolnE-1}
\epsilon = - \varepsilon_0 +  2 \thinspace \Omega_{c} \int_{k_{\mu}}^{k_{\Lambda}} dk \left. {|v_{\ell,s} |^2  \over E_{\ell,s} - \mu  +  \epsilon} \right|_{{n=-1 \atop \tau=s}}.
\end{equation}
The integrand is strongly peaked at the chemical potential. Thus for $|\epsilon| \ll |\varepsilon_0|$, the shift is

\begin{equation}   \label{eq:varSolnE-3}
{\epsilon} \approx -(\Lambda - |\mu|) \text{e}^{\varepsilon_0 / 2 g(\mu) |v_{\mu} |^2}
\end{equation}

where $g(\mu) = {\sqrt{3} \over 8 \pi   t^2 } \left| 2\mu - \lambda \right|$ is the density of states (times $\Omega_{c}$) and $ |v_{\mu} |^2 \equiv {V_{1}^2}  w_{\mu}^{2} =  {V_{1}^2 \over 2  }  \of{ 1 - {\Delta - \lambda  \over  2|\mu| + \lambda} } $ is the effective type I hybridization, at the chemical potential.

\begin{figure}
\includegraphics[width=\linewidth]{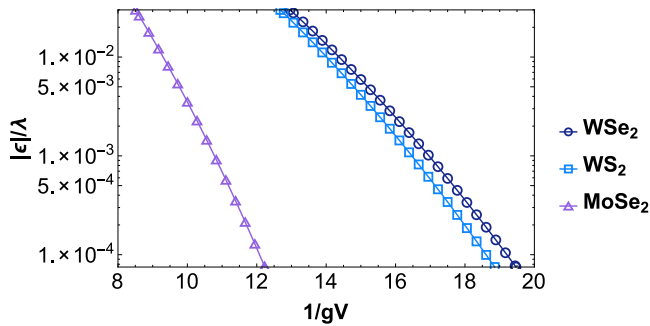}
\caption{Kondo energy scale for TMDs as a function of the inverse of the hybridization normalized to the spin splitting and density of states respectively, for the case where the chemical potential is halfway between the spin split valence bands.}
 \label{tk}
\end{figure}

For $\lambda=0$ there is no spin splitting. Thus the density of states is typically doubled compared to the spin split case studied here. Thus the Kondo temperature is lowered for large $\lambda$\cite{isaev,yanagisawa,ajikondo}. In fig. \ref{tk} we plot the Kondo energy scale $\epsilon$ as a function of the inverse hybridization strength times the density of states: $g V \equiv g(\mu) V_{1}$ with $\Lambda = \Delta$ and $\varepsilon_{0} = - \Delta/20$.  Results are shown for different TMDs when the chemical potential is halfway between the spin split valence bands, $\mu = -\Delta/2$. The larger density of states and a larger deviation away from the pole of the Bloch sphere leads to an enhanced Kondo scale for WS$_{2}$ and WSe$_{2}$ as compared to MoSe$_{2}$, revealing the mixed influence of the band and its topological character.

\subsection{Spin and Susceptibiltiy}
Since time reversal symmetry is not broken, the expectation value of the impurity spin $\langle\vect{S}\rangle$ and the electron spin $\langle\vect{\sigma}\rangle$ individually go to zero. We verify this by explicitly computing $\langle\vect{S}\rangle$ and $\langle\vect{\sigma}\rangle$. Since only the $m=0$ component of the triplet is admixed, the $x$ and $y$ components are automatically zero. The $z$ components are given by

\begin{eqnarray}   \label{eq:varMag1}
\left< \psi \right| S_z \left| \psi \right> \equiv \left< S_z \right> &=&  {\sum_{\ell ,s}}' s |B_{\ell ,s} |^2=-\left< \sigma_z \right> \\
\nonumber
&=&  {\sum_{\ell }}' ( |B_{\ell ,\uparrow} |^2 - |B_{\ell ,\downarrow} |^2 ).
\end{eqnarray}

The sum is zero due to time reversal symmetry.
The existence of the triplet component implies that the impurity is under screened. Thus we consider the expectation value of the total spin $\vect{J}^{2} = (\vect{S}+\vect{\sigma})^{2}$. Due to spin orbit coupling in the pure system the ground state $\state{\psi_{0} }$ does not have a simple singlet configuration, so the meaningful quantity is $\langle J^2 \rangle \equiv \cstate{\psi} \vect{J}^{2} \state{\psi} - \cstate{\psi_{0}} \vect{\sigma}^{2} \state{\psi_{0}}$.  Defining also $\delta\theta_{k} \equiv \theta_{k,+,\uparrow} - \theta_{k,+,\downarrow}$, the difference in polar angle on the Bloch sphere of opposite spin states,

\begin{eqnarray}
\langle J^2 \rangle &=& 2 {\sum_{\tau, \vect{k} = k_{\mu}}^{k_{\Lambda}} } \cos{{ \delta\theta_{k}\over 2} } \left[ |p_{-1,\vect{k},\tau}|^{2} \of{\cos{{ \delta\theta_{k}\over 2} }  - 1 } \right.		\nonumber \\
&& + \left. |t_{-1,\vect{k},\tau}|^{2} \of{\cos{{ \delta\theta_{k}\over 2} }  + 1 } \right]		\label{eq:TotalJ-hole}  \\
&\approx&	  \cos^{2}{{ \delta\theta_{\mu}\over 2}}  .		\label{eq:TotalJ-apx}
\end{eqnarray}

\begin{figure}
\includegraphics[width=\linewidth]{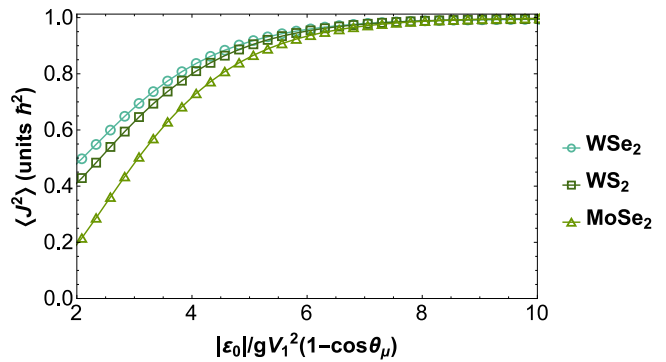}
\caption{The expectation value of $J^{2}$ is plotted as a function of the effective coupling that controls the Kondo scale (see eq. (\ref{eq:varSolnE-3})).}
 \label{j2}
\end{figure}

In fig. \ref{j2} we plot the variation of $\langle J^{2} \rangle$ as a function of the exponent on the RHS in eq. (\ref{eq:varSolnE-3}). For weak hybridization, the resonance is an equal mixture of singlet and triplet and $\langle J^{2} \rangle \approx \hbar^{2}$. The interacting system remains nonmagnetic, as revealed by $\langle \vect{J} \rangle=0$, but fluctuations give $\langle {J}^{2} \rangle \neq 0$. Note that only the contribution from the Kondo resonance excluding the Fermi sea is given by eq. (\ref{eq:TotalJ-hole}). As the hybridization gets larger so does the width of the Kondo resonance leading to a decrease in $J^{2}$. Since the spin splitting in WS$_{2}$ and WSe$_{2}$ is large compared to MoSe$_{2}$, the deviation away from $\hbar^{2}$ occurs for a larger value of $V_{1}$ for the former two. 
 
Concluding this section we consider the magnetic susceptibility.  Note that the conservation of the $z$-component of spin yields an anisotropic susceptibility.  For the ground state considered here we focus on the out of plane response. To do so we couple the magnetic field to the impurity spin which further splits the energies: $\varepsilon_{\ell,s} \to \overline{\varepsilon}_{\ell,s} = \varepsilon_{\ell,s} + \mu_{0} h_{z} s$. The only nonzero component of the susceptibility tensor is $\chi^{imp} \equiv \chi^{imp}_{zz} = { d \over d h_{z}} \mu_{0} \langle S_{z} \rangle$. The zero field  value is

\begin{equation}   \label{eq:gndImpSus}
\chi^{imp} |_{\vect{h}=0} = {2 \mu_{0}^{2} b_{0}^{2} \over {N_{M}}} {\sum_{\ell,s}}' {|v_{\ell,s} |^2  \over | \varepsilon_{\ell,s}  +  \epsilon |^{3} } \sim |\epsilon|^{-1} .
\end{equation}

Thus a finite spin orbit coupling reduces the Kondo energy scale enhancing the susceptibility.


\section{Numerical Renormalization Group}     \label{sec:nrg}

An important property of TMDs is the valley selective coupling of circularly polarized light. This is allowed because of the topological nature of the low-lying states in monolayer TMDs which allows for optical transitions between two bands of $d$ character. The transition rates for incident circular polarization are valley discriminating in that right handed polarization couples predominantly to one valley and left to the other\cite{xiaodi}. For a hole doped system this suggests that the Fermi surface in the two valleys are at different energies when irradiated. Therefore the case of a magnetic impurity in a Fermi sea with effectively offset spin up/down chemical potentials is of particular interest. Instead of using the variational wave-function technique which requires an ansatz for the possible ground state, we turn to an unbiased approach. In this section we analyze the same system, including both zero and finite offset of chemical potentials, using Wilson's numerical renormalization group (NRG) method \cite{wilsonnrg}. 

We are interested in the ground state of the Anderson Hamiltonian in hole-doped monolayer TMDs with the emphasis on  tuning the properties of the many-body Kondo bound state by application of circularly polarized light. As stated above, valley-selectivity translates to spin-selectivity. Thus under a steady fluence a Fermi sea  is generated with unequal chemical potentials for spin up/down. The effective difference in chemical potentials is denoted $\delta$, and the midpoint of the potentials $\mu$.

We begin with a brief description of the standard procedure used to investigate the NRG flow adapted to the TMD system. The Anderson problem is first mapped to semi-inifnite linear chain with nearest neighbor hopping starting with the impurity at one end. Tracking the flow towards the fixed point Hamiltonian enables the computation of thermodynamic properties such as the impurity occupation, entropy, moment, and susceptibility. Additionally the Kondo temperature as well as the impurity spectral functions are determined. We perform the analysis first for the case of equal chemical potential in both valleys, where the results are shown to agree with the variational approach, before exploring the case with finite $\delta$.

Details for the following transformations, diagonalization, and results are presented in Appendix \ref{app:nrg}.

\subsection{Generic Setup}     \label{sec:nrgsetup}

Since only the upper valence bands cross the chemical potential we include only one band. The other spin-split bands are filled and of $\mathcal{O}(\lambda)$ away in energy to affect the low energy physics occurring at the Kondo scale. This simplified picture is most applicable when the spin-splitting is largest, so the most appropriate materials are WSe$_{2}$ and WS$_{2}$. In that case, and due to the fermionic constraint $n_{f,s}\of{n_{f,s}-1} = 0$, the Anderson Hamiltonian may be written as 

\bea
H' &=& H - \sum_{s=\uparrow,\downarrow} \mu_{s} N_{s}  \nonumber \\
&=& \sum_{\vect{k},s} \of{ E_{k} - \mu_{s} } c_{\vect{k},s}\dag  c_{\vect{k},s}  + \sum_{\vect{k},s} \of{ v_{\vect{k}}^{\alpha}  c_{\vect{k},s}\dag  f_{s}  +  h.c. }  \nonumber	\\
&& + \sum_{s} \of{\varepsilon_{0} - \mu_{s}} f_{s}\dag f_{s}  + U n_{f,\uparrow} n_{f,\downarrow}  \nonumber  \\
&=&  H_{m} + H_{V}^{\alpha} + \sum_{s} \of{\varepsilon_{0} - \mu_{s} + {U \over 2} } f_{s}\dag f_{s}	\nonumber	\\
&& +   {U \over 2} \of{\sum_{s}n_{f,s} - 1 }^{2}   -  {U \over 2}
\label{eq:nrgH1}
\eea
where the number operators are $N_{s} = n_{f,s} + \sum_{\vect{k}} c_{\vect{k},s}\dag  c_{\vect{k},s}$, and $H_{m}$ and $H_{V}^{\alpha}$ are the material (TMD) and hybridization Hamiltonians as in eqs. (\ref{eq:metalHmod2}) and (\ref{eq:type1Hyb}), respectively. The remaining terms (aside from $U/2$) are now collected to form $H_{imp}$. The chemical potentials are $\mu_{s} = \mu + s \delta / 2$ so that the difference is $\delta$. The low-energy model leading to $H_{m}$ is only valid within a cutoff, $|E_{k}| < \Lambda$ of the order of the gap $\Delta$. Offsetting about the hole doped chemical potential, we set $D=\Lambda+\mu$ as the effective cutoff and we note the top of the band $e_{0} = E(0) - \mu$ so that $-D < E(k)-\mu < e_{0}$.

As in the previous sections, the form of the hybridization  corresponds to the case where the impurity shares a metal (i.e. W, Mo) site with type $\alpha = 1 \text{ or } 2$ orbital. The operators appearing in $H_{m}$ and $H_{V}^{\alpha}$ above refer only to the upper valence band, with labels $n=-$ and $\tau=s$.

We follow the standard procedure\cite{kmww1} and construct effective states labeled with energy $\varepsilon$ relative to the chemical potentials. Specifically, we use the quasi-angular momentum (QAM) operators as in eq. (\ref{eq:type1Hybalt}) and use the energy density of states $g(\varepsilon)$ to keep track of the measure. The continuum limit is effectively achieved, and only states hybridizing with the impurity, i.e. those with QAM $\nu=0$ for type I or $\nu=-\tau$ for type II, are included. The remaining states, those with QAM labels not appearing in $H_{V}^{\alpha}$, are inert just like the lower valence bands --- all such states are ignored here as they will not be affected by the presence of the impurity.

The approximation of NRG originates from logarithmic discretization, which introduces a parameter\footnote{The notation here is different from that of ref.s \onlinecite{kmww1,costinrg} and others since their symbol, $\Lambda$, is used here for the model energy cutoff.} $R>1$. The transformation is exact for $R\to1$. The choice $R=3$, used in the analysis below, is standard for implementation which strikes a balance between speed and accuracy.

Projecting to form a semi-infinite chain with nearest-neighbor hopping for the TMD electrons, the Hamiltonian is   
\bea
{H' \over D} &\approx&  \sum_{m=0}^{\infty} \sum_{s} \of{ \epsilon_{m,s} d_{m,s}\dag d_{m,s} + t_{m,s} \of{d_{m+1,s}\dag d_{m,s} + h.c.} }   \nonumber \\
&& + \sqrt{\Gamma (1+e_{0}/D) \over \pi D} \sum_{s} \of{d_{0,s}\dag f_{s} + h.c.} + {H_{imp}  \over D}
\label{eq:nrgH4}
\eea
where $\Gamma = \pi g(\mu) |v_{\mu}^{\alpha}|^{2}$, and the site energies/hoppings ($\epsilon_{m,s}$ and $t_{m,s}$) are determined numerically. 

The NRG scheme uses a rescaled Hamiltonian $\overline{H}_{M}$ for a finite number of sites $M$. The hopping parameter is known to scale as $R^{-m/2}$. Therefore the Hamiltonian is rescaled $\propto R^{M/2}$ to ensure that each additional site added to the chain only increases energy by $\mathcal{O}(1)$\cite{wilsonnrg,kmww1}. Specifically, the Hamiltonian (\ref{eq:nrgH4}) is recovered from the rescaled finite version with 
\be
{H' \over D} = \lim_{M\to\infty} \left[ {1 + R^{-1}  \over 2} R^{-(M-1)/2} \overline{H}_{M}  \right]	.
\label{eq:nrgHlim}
\ee
To solution proceeds as follows. First the system is diagonalized starting with $\overline{H}_{0}$ for the impurity and one site representing the chain. Then additional chain sites are added one-by-one, diagonalizing after each new Hamiltonian is built. To prevent the exponential growth of Fock space, the spectrum is truncated so that only the lowest-energy states are kept. In this way, the solution for the infinite system is obtained by solving finite systems in the limit of large $M$; in practice, for $R=3$, a chain of at least $M=30$ sites is needed for good convergence.

\subsection{System in Equilibrium ($\delta=0$)}		

\subsubsection{Thermodynamics}\label{sec:thermo}

The solution of the procedure above is the starting point to compute thermodynamic and spectroscopic properties of the Kondo resonance. The main object in thermodynamics, the partition function, is 
\bea
Z(T) &=& \Tr \exp\of{-\beta H'} = \lim_{M\to\infty} \Tr \exp\of{-\overline{\beta}_{M} \overline{H}_{M} }	\nonumber	\\
&\equiv& \lim_{M\to\infty} Z_{M}(T)
\label{eq:nrgZ}
\eea
where $\beta=1/k_{B}T$ is the usual inverse temperature while $\overline{\beta}_{M} = ((1 + R^{-1}) / 2) R^{-(M-1)/2} D/k_{B}T$ is a rescaled dimensionless version. Since we have used $R=3$ we choose to fix $\overline{\beta}_{M} = 1/2$, defining a set of exponentially decreasing temperatures corresponding to the chain length.

\begin{figure}[ht]
\includegraphics[width=\tabfigwidth]{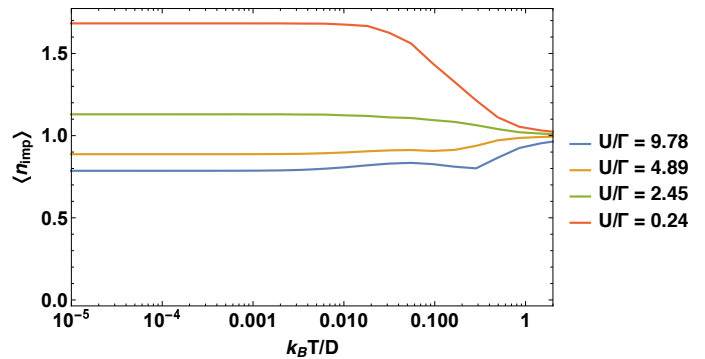}   
\caption{Impurity occupation in the unexcited case, $\delta=0$, for several choices of $U$ at fixed $\Gamma$ (see sec. \ref{sec:numres} for parameter details). The occupation is diminished or enhanced from $1$ due to the asymmetry of the impurity level about the chemical potential.}
\label{fig:1imp}
\end{figure}

Since we are primarily interested in the behavior of the impurity we focus on the occupations of each spin,
\bea
\langle n_{f,s} \rangle &=& \Tr \left[ f\dag_{s}f_{s} \exp\of{-\overline{\beta}_{M} \overline{H}_{M} } \right] /Z_{M}	\nonumber	\\
&=& \sum_{q} \cstate{q} f\dag_{s}f_{s} \state{q} \exp\of{-\overline{\beta}_{M} \overline{E}_{q} } / Z_{M}	
\label{eq:impocc}
\eea
where we have introduced eigenstates $\state{q}$ with energies $\overline{E}_{q}$ (in the space up to site $M$). In order to calculate such objects, one must keep track of the matrix elements $\cstate{q} f\dag_{s}f_{s} \state{q}$ projected to the eigenspace of each successively longer chain. With matrix elements and energies in hand, quantities like (\ref{eq:impocc}) are straightforward to calculate, giving impurity occupation and spin values $\langle n_{imp} \rangle = \sum_{s} \langle n_{f,s} \rangle $ and $\langle s_{z} \rangle = (\langle n_{f,\uparrow} \rangle - \langle n_{f,\downarrow} \rangle) / 2$. To display the typical behavior in our case, we show in fig. \ref{fig:1imp} the impurity occupation for WSe$_{2}$ with equal chemical potentials (the spin is omitted because it is identically zero in this case).

\begin{figure}[ht]
\begin{tabular}{ c }
\includegraphics[width=\tabfigwidth]{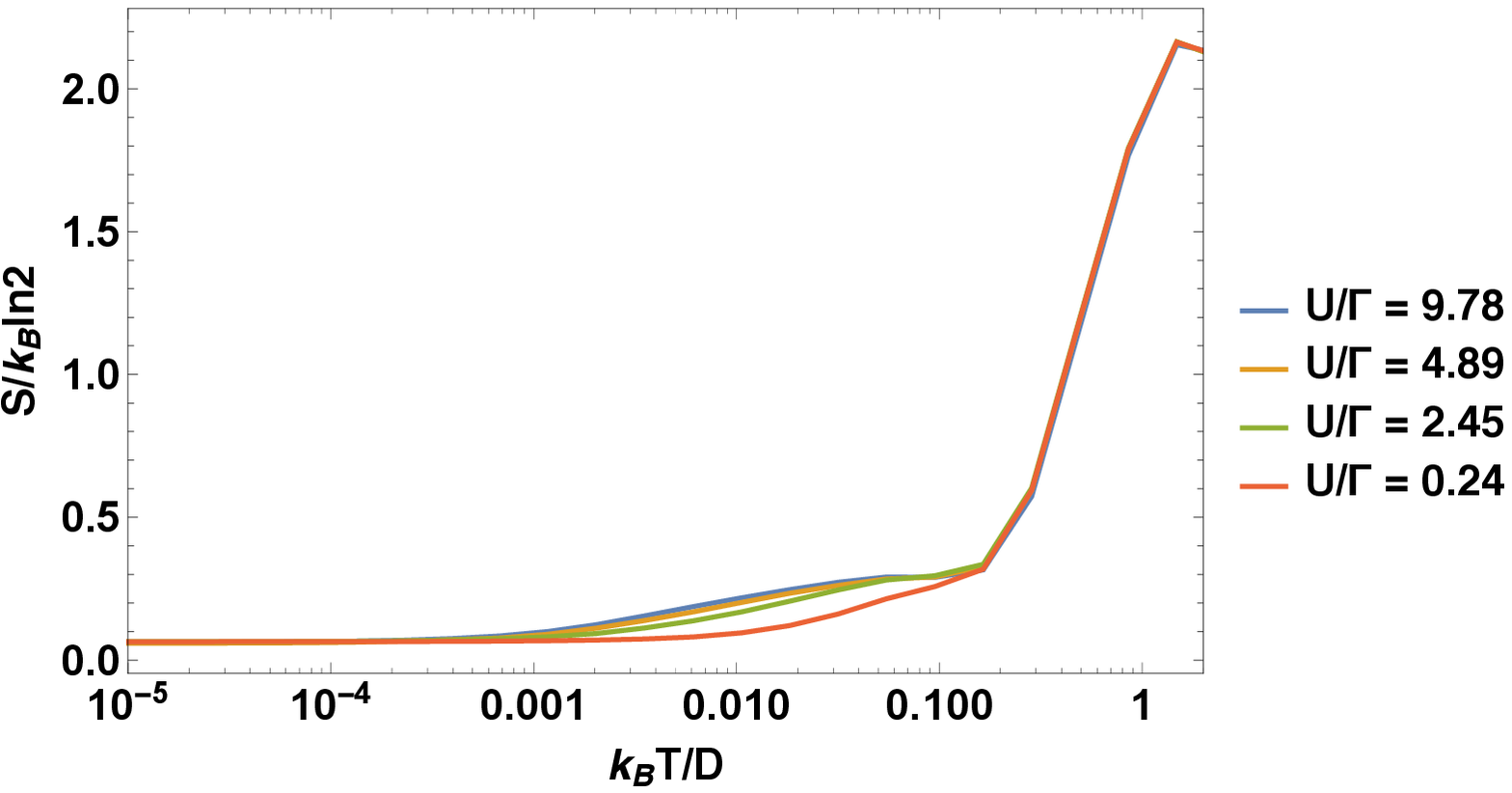}    \\
\hline
     \includegraphics[width=\tabfigwidth]{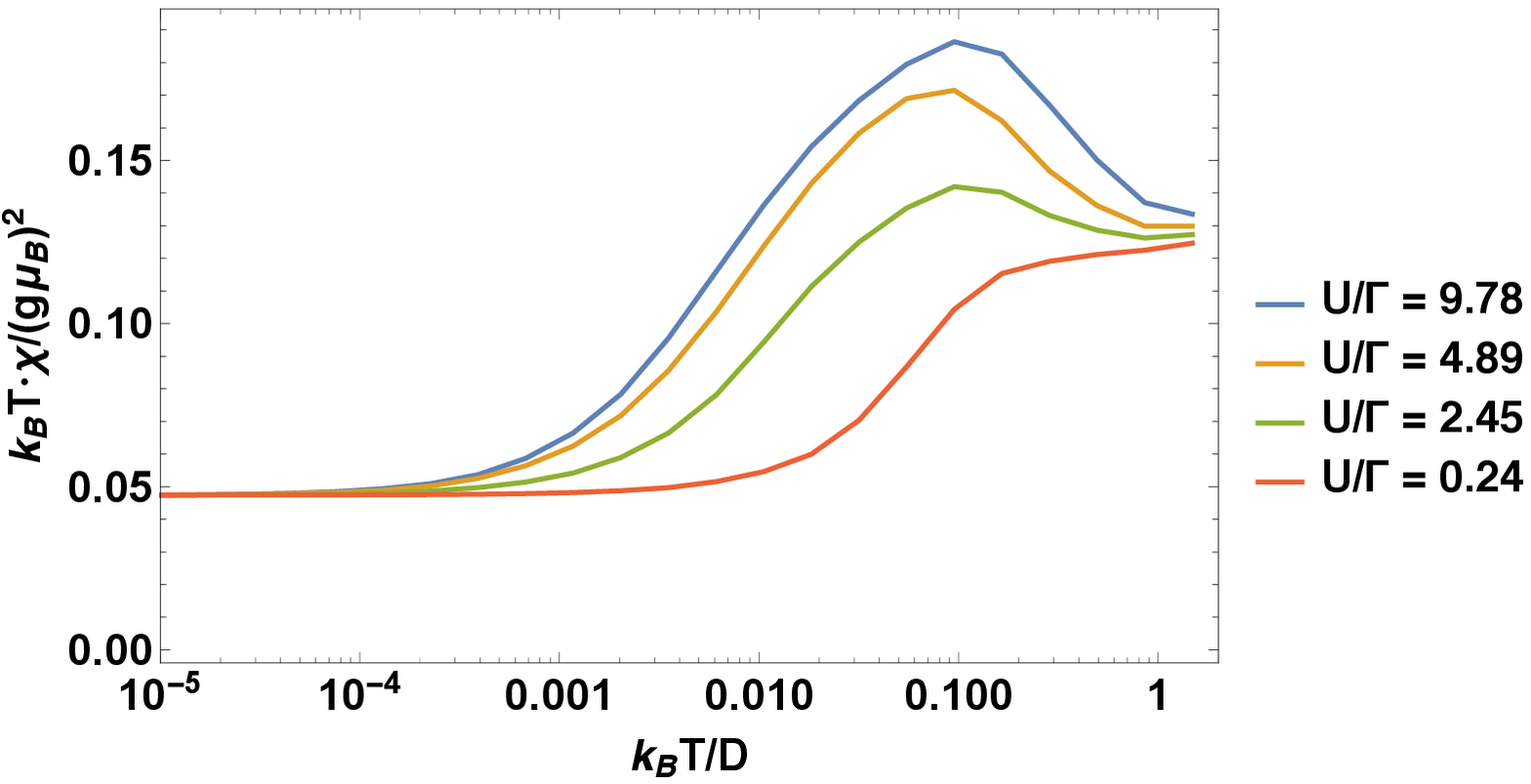}
\end{tabular}
\caption{Impurity entropy (top) measured in units of $k_{B}\ln2$ and susceptibility (bottom) in units of $(g\mu_{B})^{2}/k_{B}T$, in the unexcited case. The trends are typical (as in fig. 6 of ref. \onlinecite{costinrg}), except for the asymptotic values. }
\label{fig:1thermo}
\end{figure}

System properties, like entropy and magnetic susceptibility, are calculated using the standard precedure; impurity contributions are determined by subtracting the result for the chain only (i.e. $H_{V} \to 0$ , $H_{imp} \to 0$) from the result for the full system with the impurity\cite{bcpnrg,wilsonnrg}.
The impurity entropy and susceptibility are shown in fig. \ref{fig:1thermo} for WSe$_{2}$, revealing rather typical behavior for the Anderson model within NRG, except that we have a nonzero value of $T\chi$ as $T\to0$ suggesting the persistence of a local moment.

\begin{figure}[ht]
\includegraphics[width=\figwidth]{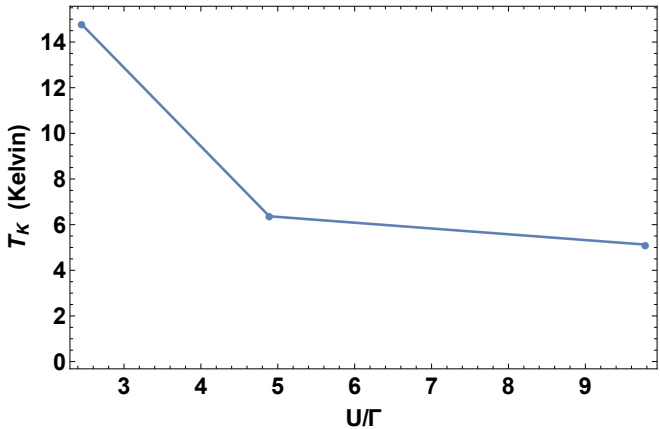}    
\caption{The Kondo temperature as given by Wilson's relation for the unexcited case of WSe$_{2}$ at a few $U/\Gamma$. The trend of increasing $T_K$ at decreasing $U/\Gamma$ is due to the proximity to the strong-coupling fixed point in parameter space.}
\label{fig:1tk}
\end{figure}

In addition to providing information about the formation or persistence of a local moment, we follow Wilson\cite{wilsonnrg} to estimate the Kondo temperature scale from the susceptibility curve. The key insight is that $k_{B}T\chi (T)$ is an universal function of $T/T_{K}$. The behavior is Curie-Weiss-like near $T_{K}$ and the best fit to the perturbative result for the susceptibility at $T_{K}$ is $k_{B}T_{K}\chi (T_{K})/(g\mu_{B})^{2} \approx 0.0567$. Shown in fig. \ref{fig:1tk} is the temperature obtained in this way for a few values of $U/\Gamma$. 

Details of the NRG flow diagrams (as in fig. 8 of ref. \onlinecite{kmww1}), which reveal the fixed points and their stability, are necessary to understand the increasing temperature with decreasing $U$. The final configuration for any starting point is the strong-coupling ($U\to0, \Gamma \to \infty$) fixed point, at which the impurity is fully bound. For small $U/\Gamma$, a local moment is never formed (i.e. the flow stays away from the local moment fixed point) and the impurity is easily bound so the Kondo temperature is higher. On the other hand, large $U/\Gamma$ places the system very near to the local moment fixed point so that the flow initially gives a local moment, and only after more iterations does the system begin to approach the strong-coupling fixed point, so the Kondo temperature is lower. This picture is not just useful for the temperature scale; the same description from the flow diagram also nicely explains the susceptibility and how it changes with various $U/\Gamma$ as seen in fig. \ref{fig:1thermo}, the presence and size of the bump at moderate temperatures revealing the formation of a local moment.

\begin{figure}[ht]
\begin{tabular}{ c }
\includegraphics[width=\tabfigwidth]{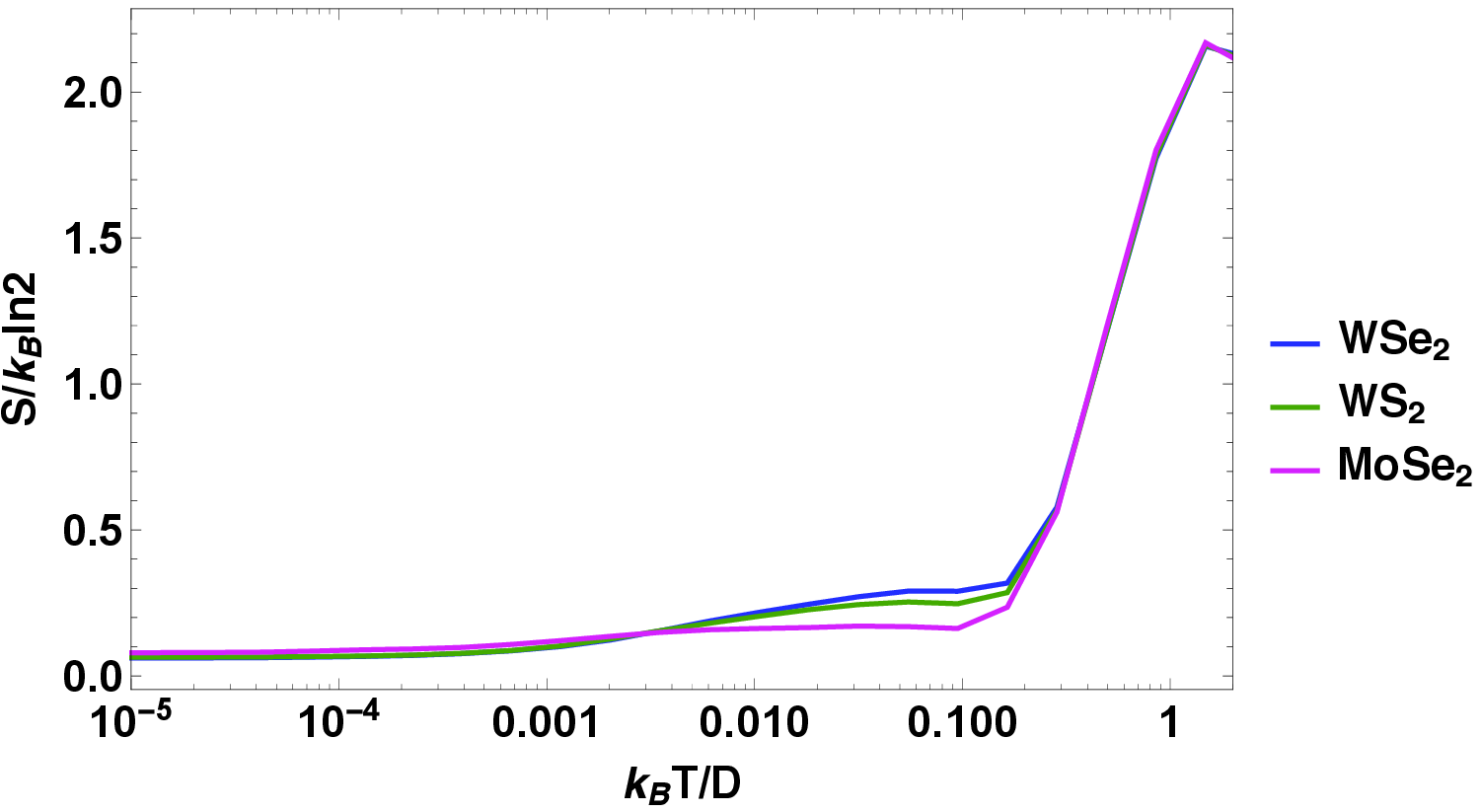}    \\
\hline
     \includegraphics[width=\tabfigwidth]{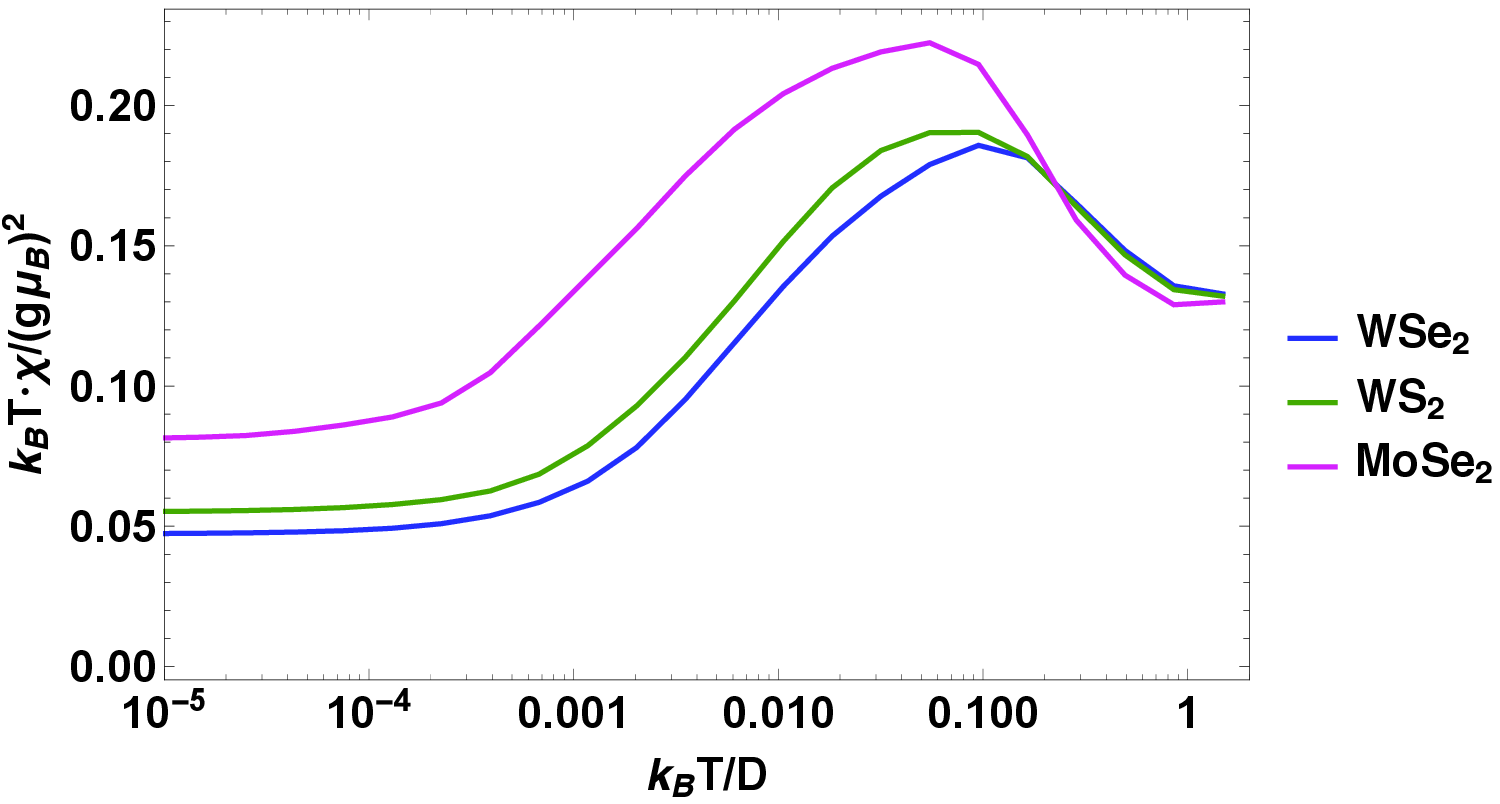}
\end{tabular}
\caption{Impurity entropy (top) measured and susceptibility (bottom) for three monolayer TMD materials. They are all in the unexcited case, and with fixed $U/\Gamma = 9$. The trends are explained in the same way as for fig. \ref{fig:1thermo}. }
\label{fig:matthermo}
\end{figure}

To compare across materials, we present in fig. \ref{fig:matthermo} the entropy and susceptibility for three TMDs (recall MoS$_{2}$ is excluded). Despite the constant ratio $U/\Gamma=9$ the materials look like they have local moment regimes of varying intensity. This is due to the topological content captured by the Bloch angles appearing in the effective coupling to the impurity, i.e. the $w=\sin(\theta_{\mu}/2)$ of eq.s (\ref{eq:type1Hyb}) and (\ref{eq:type1Hybalt}). The angle $\theta$ approaches zero as the chemical potential approaches the lower valence band top, the distance from which is given by the magnitude of the spin-orbit coupling $\lambda$. Thus the Tungsten compounds with larger $\lambda$ start off with larger $\Gamma$ and their NRG trajectories form relatively weaker local moments, quickly going to the strong-coupling fixed point. On the other hand, MoSe$_{2}$ has the smallest spin splitting and hence the smallest deviation of the Bloch angle from the pole so its NRG trajectory begins close to the free orbital fixed point ($U=\Gamma=0$), leading quickly to a strong local moment, and very slowly to the strong-coupling point. This variation among the three materials and the explanation related to the Bloch angles are consistent with the discussion in sec. \ref{sec:varpar}, following eq. (\ref{eq:varSolnE-3}).

\subsubsection{Spectral Function}

The impurity spectral function for each spin provides information on one particle properties such as the spin polarization and impurity occupation. Since we are interested in ground-state properties, only the $T=0$ spectral function is required. It is given by \cite{bcpnrg,bullanrg}

\bea
A_{s}(\omega) &=& { 1 \over Z(0) } \sum_{q} \left[ \left| \cstate{q} f_{s} \state{0}  \right|^{2} \delta\of{\omega + (E_{q} - E_{0})}  \right.	\nonumber	\\
&& + \left.  \left| \cstate{0} f_{s} \state{q}  \right|^{2} \delta\of{\omega - (E_{q} - E_{0})}  \right]
\label{eq:spectdef}
\eea
where the energies $E_{q}$ are in the space of the full system Hamiltonian $H'$ as in (\ref{eq:nrgH4}).  

To construct smooth spectral functions within NRG\cite{bullanrg,costinrg} we implement the conventional method where the $\delta$-peaks are smoothened  using Gaussians. While the method underestimates the spin polarization when time reversal symmetry is broken, it provides qualitative predictions and a lower bound for expected behavior of the resonance.

\begin{figure}[ht]
\includegraphics[width=\tabfigwidth]{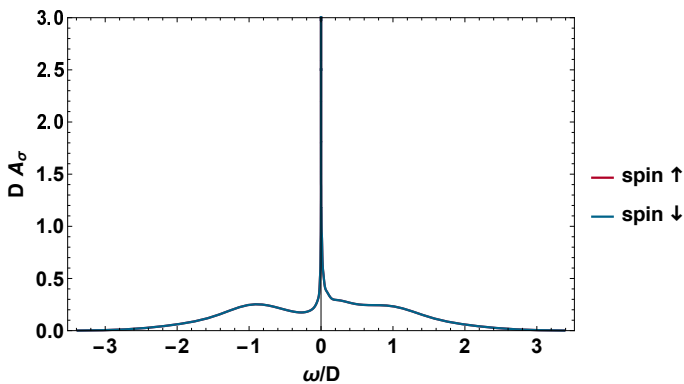}    
\caption{Impurity spectral functions for the unexcited case, with fixed $U/\Gamma\approx4.9$. The spin up/down results overlap due to preserved time-reversal. They are consistent with conventional models\cite{hewson}, as expected from the metal-like structure of our model and choice of parameters.}
\label{fig:1spec}
\end{figure}

The conventional construction and Gaussian broadening lead to final spectral functions, representatives shown in fig. \ref{fig:1spec} for the hole doped TMD case with equal chemical potentials in WSe$_{2}$. The asymmetry in shape is related to asymmetry of the impurity levels $\varepsilon_{0}$ and $\varepsilon_{0} + U$ about the chemical potential.

\subsection{Valley Asymmetry  ($\delta\neq0$)}	  \label{sec:numres}

We turn to the case when one has slightly different chemical potentials in the two valleys/spins due to optical excitation.
The effective polarization is denoted $\delta=\mu_{\uparrow}-\mu_{\downarrow}$, and the midpoint of the potentials is $\mu=(\mu_{\uparrow}+\mu_{\downarrow})/2$.

We choose parameters so as to compare with the variational results.  The midpoint is fixed at $\mu=-\Delta/2$ (midway between the spin-spit valence bands) and the impurity is set just below the chemical potential $\varepsilon_{0} = -\Delta/20$, placing the model in a ``mixed-valence" regime\cite{hewson}. Several polarization cases are included: $\delta=0$, $\delta=\pm\lambda/10$, and $\delta=\pm\lambda/5$ (see eq. (\ref{eq:metalHamil}) for a reminder of the band parameters).  Note that  $\delta/2 < |\varepsilon_{0}|$ to ensure that the impurity level is below both chemical potentials; the largest splitting parameter is $\lambda\sim\Delta/6$ so the condition is satisfied, using the largest value of $\delta$ above, with $1/60 < 1/20$. Also, using an absolute cutoff energy scale $\Lambda=\Delta$, the effective cutoff relative to $\mu$ is $D=\Delta/2$. To neglect the hybridization of the lower (filled) valence bands, we choose the material with the largest spin-splitting: WSe$_{2}$. The width $\Gamma$ depends on the hybridization strength set to $V_{1}=0.56 t$ for type $\alpha=1$ coupling, i.e. approximately half of the hopping energy, leading to $\Gamma/D \approx 0.051$. The Coulomb repulsion is fixed at at $U=\Delta/8=D/4$, giving $U/\Gamma \approx 4.9$, unless otherwise stated.

Before we discuss the nature of the resonance, it is important to point out a vital difference between this case and other ``asymmetrical" cases studied in the context of Kondo phenomena. A typical ``asymmetrical" model\cite{kmww2} has the impurity level set away from $-U/2$ so that the bilinear impurity term survives in (\ref{eq:nrgH1}). Here we also have the chemical potential close to the band maxima so that the energy interval is ${[-D,e_{0}]}$ with $e_{0}<D$, giving the model site energies $\epsilon_{m}\neq0$ as well.

\begin{figure}[ht]
\begin{tabular}{ c }
\includegraphics[width=\tabfigwidth]{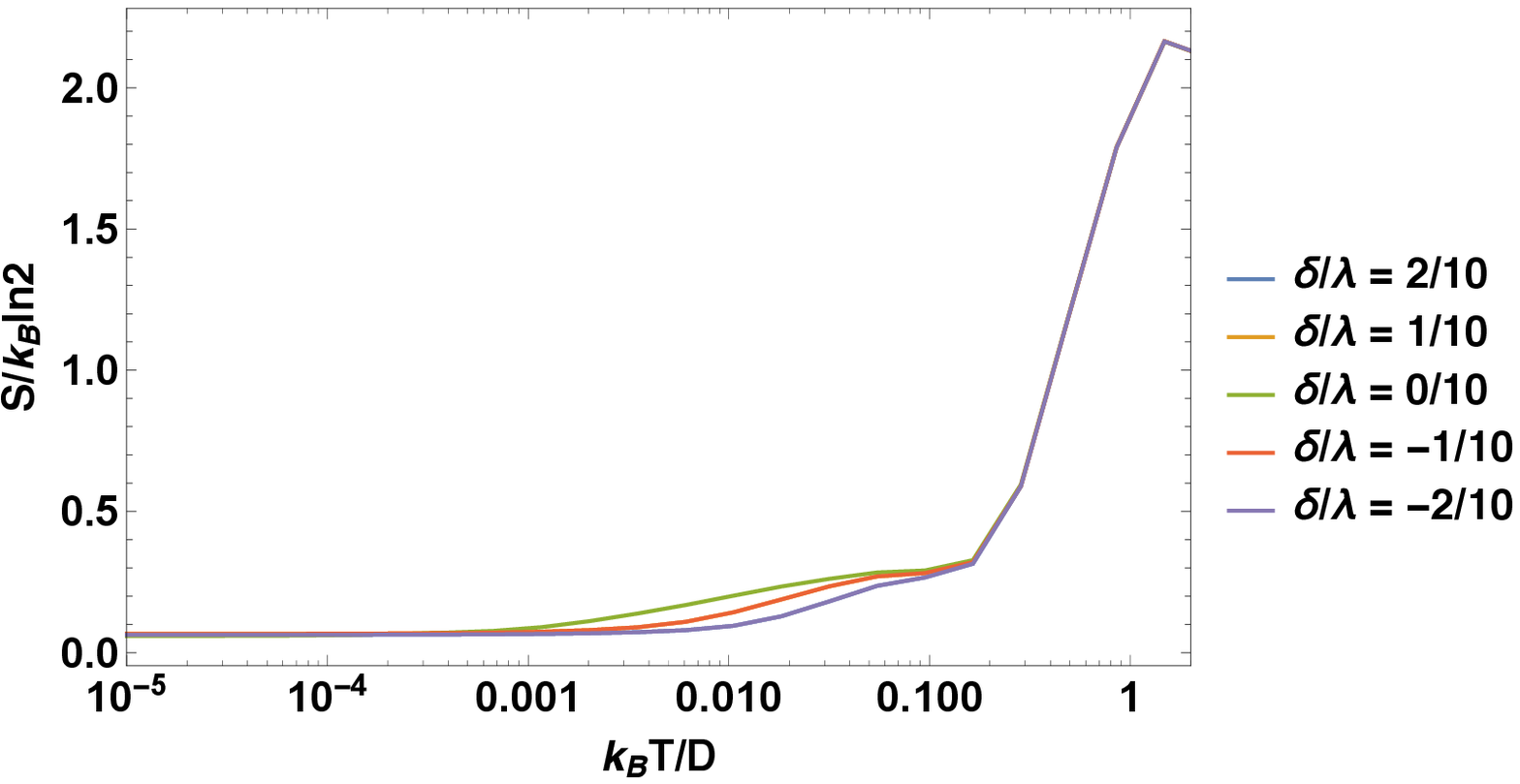}    \\
\hline
     \includegraphics[width=\tabfigwidth]{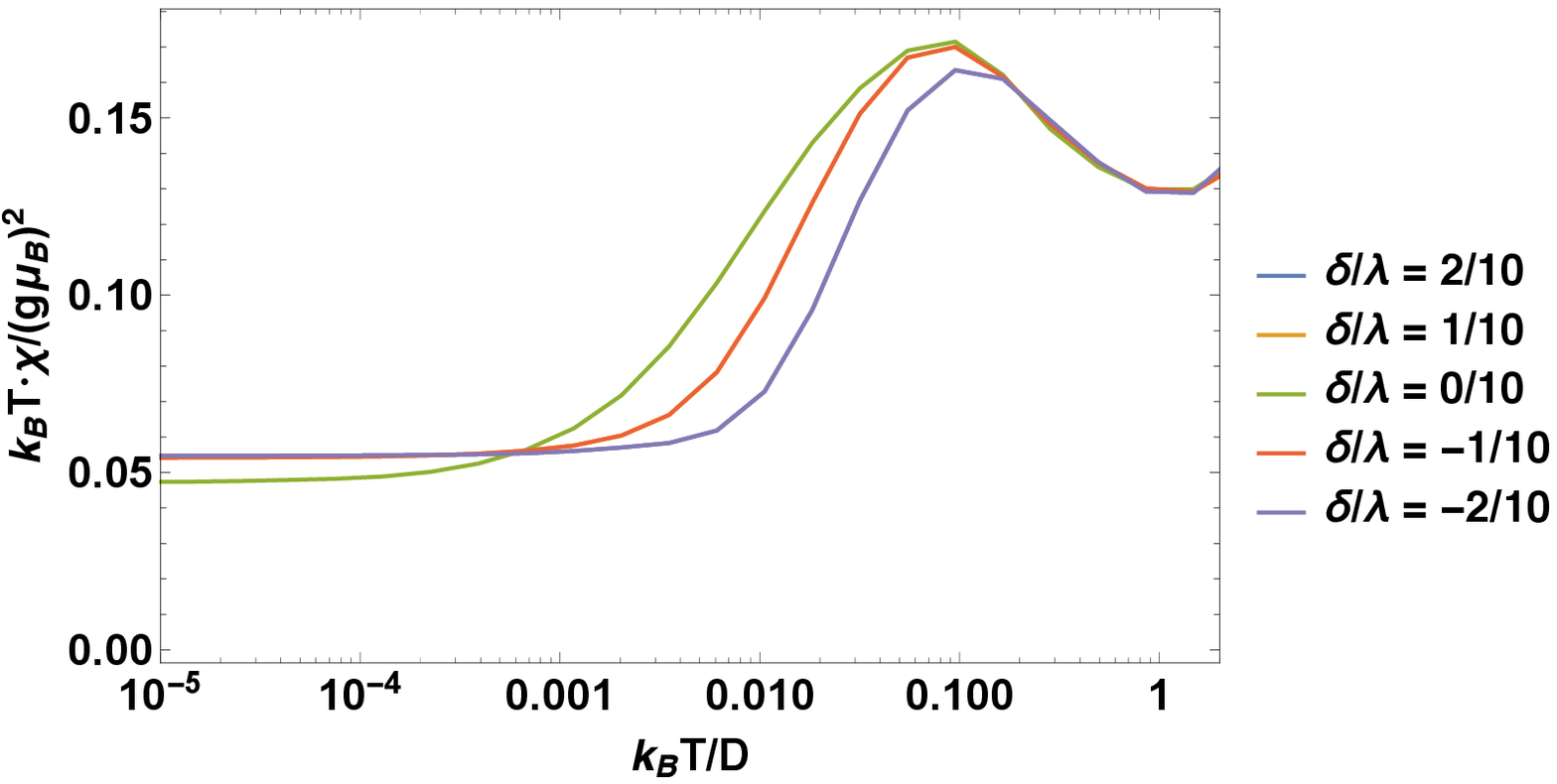}
\end{tabular}
\caption{Impurity entropy and susceptibility in the excited system for various offset $\delta$. The numerical values are as described at the beginning of this section. The rightward evolution reveals increasing $T_K$.}
\label{fig:2thermo}
\end{figure}

\begin{figure}[ht]
\includegraphics[width=\figwidth]{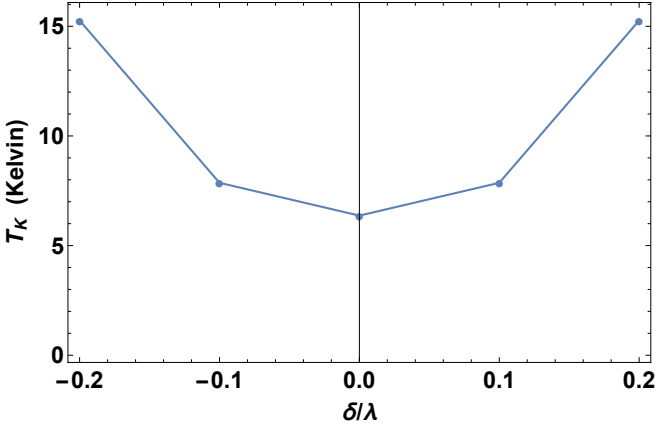}    
\caption{The Kondo temperature in the excited case at a few $\delta$. The increasing values away from $\delta=0$ suggest effectively larger $\Gamma$ with increasing magnitude of $\delta$, pushing the starting point closer to the strong-coupling fixed point.}
\label{fig:2tk}
\end{figure}

The entropy and susceptibility, fig. \ref{fig:2thermo}, show little difference compared to those of the previous fig. \ref{fig:1thermo}. The susceptibility lands higher as the temperature is decreased, corresponding to a larger moment surviving at $T=0$. Also, both suggest an increasing $T_K$ with $\delta$ since the asymptotic values are approached more quickly; indeed that is what we observe in fig. \ref{fig:2tk}. The larger moment is expected due to the the increase in degree of broken time reversal symmetry, but the larger Kondo temperature is surprising. The evidence suggests that, due to the difference in energies for spin up/down, the system is rapidly pushed in the direction of the inevitable moment formation and $T_K$ is increased.

\begin{figure}[ht]
\begin{tabular}{ c }
\includegraphics[width=\tabfigwidth]{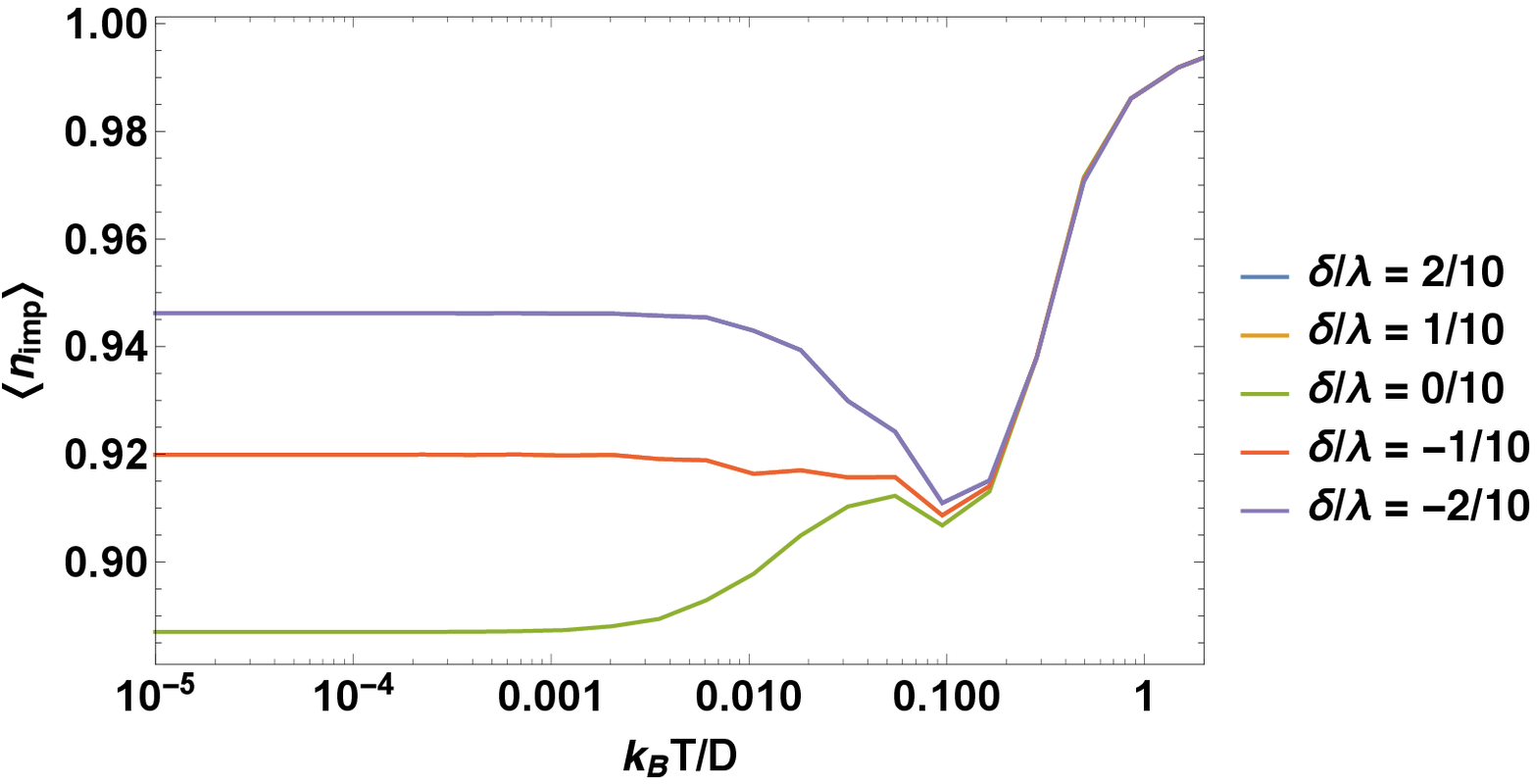}    \\	
\hline     
\includegraphics[width=\tabfigwidth]{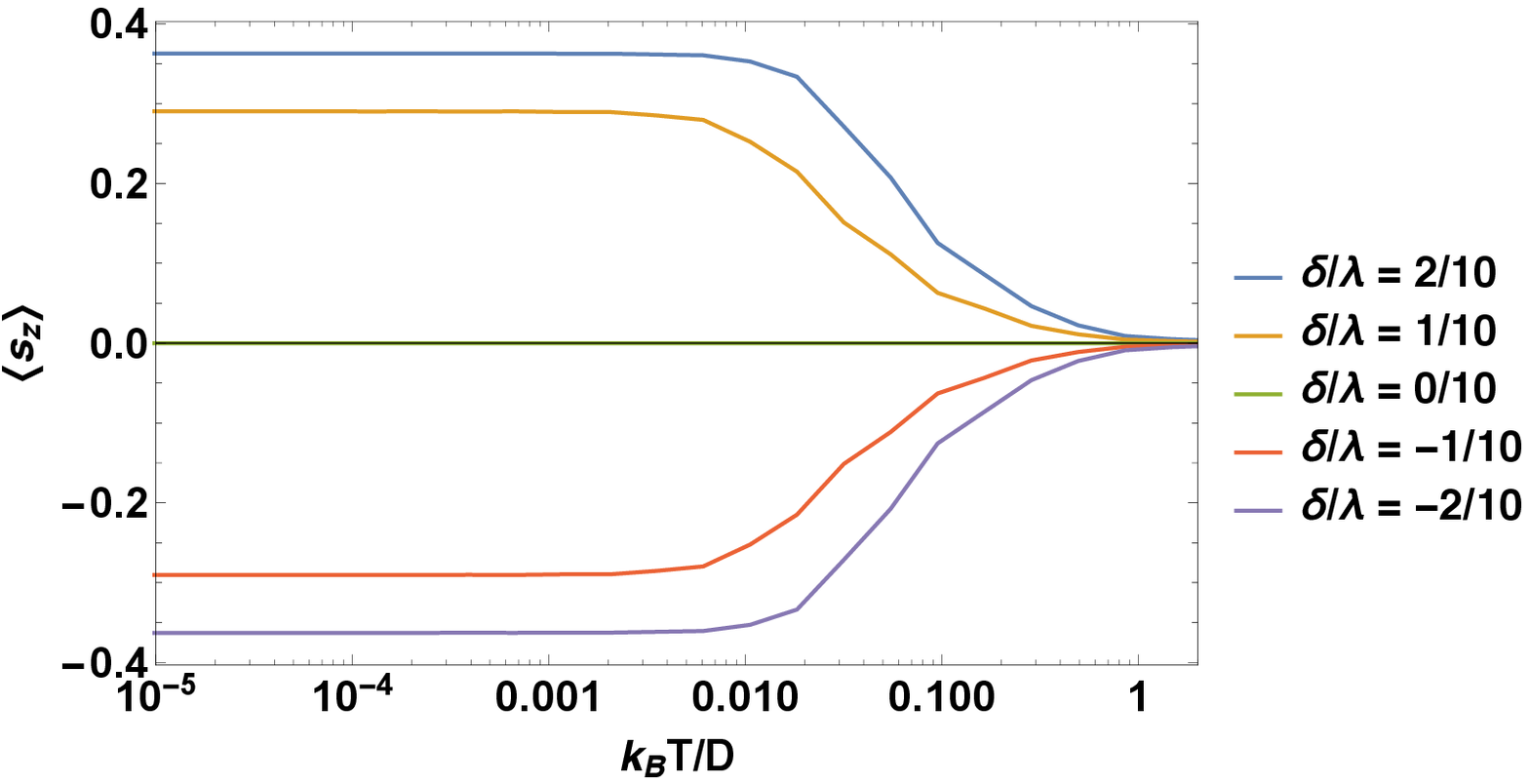} 
\end{tabular}
\caption{Impurity occupation (top) and spin (bottom). The nonzero offset polarizes the spin as expected. The occupation is increased with larger $\delta$ because one chemical potential becomes more distant from the impurity level.}
\label{fig:2imp}
\end{figure}

Examining impurity properties further, the occupation $\avg{n_{imp}}$ and spin $\avg{s_{z}}$ are presented in fig. \ref{fig:2imp}.  
The spin is polarized overall with sign and magnitude reflecting the difference $\delta$. As for the occupation, the value is increased because one of the impurity levels (up or down) is effectively deeper below its chemical potential, increasing the occupation of that spin, while the other is shallower but still remains below its chemical potential, preventing its value from decreasing too much. The bump at moderate temperatures, which shrinks with increasing $\delta$, aligns closely with the bump seen in susceptibility which marks the point at which the system begins to depart the local moment regime and moves toward the strong-coupling fixed point. Similarly, the average occupation drops as the local moment is formed since double occupation is not present (the effective $U/\Gamma$ becomes very large), while the approach to strong-coupling allows for a growth of occupation (now $U/\Gamma \to 0$) if the effective depth of the impurity is large enough.

\begin{figure}[ht]
\includegraphics[width=\figwidth]{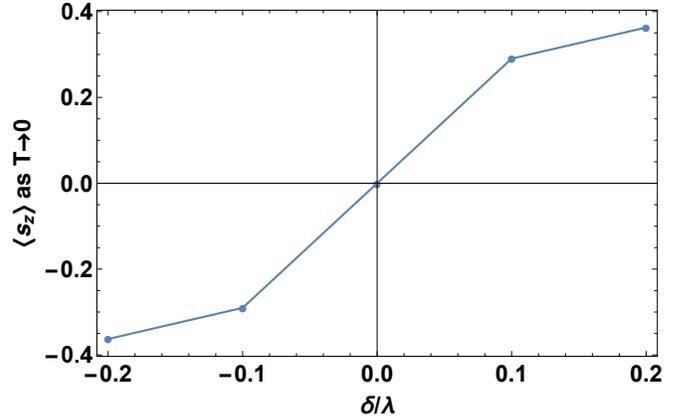}   
\caption{The impurity spin at $T\approx0$ shows some familiar dependence on $\delta$. This result is equivalent to the shift in total spin $J_{z}$, the contribution from the pure TMD having been subtracted.}
\label{fig:2gspin}
\end{figure}

To better understand the spin at $T=0$, we show in fig. \ref{fig:2gspin} the impurity spin at the lowest available temperature plotted against the offset $\delta$. Our offset choice results in only five points total, but the trend is reminiscent of the familiar hyperbolic tangent for a paramagnetic system with $\delta$ playing the role of magnetic field.

\begin{figure}[ht]
\includegraphics[width=\tabfigwidth]{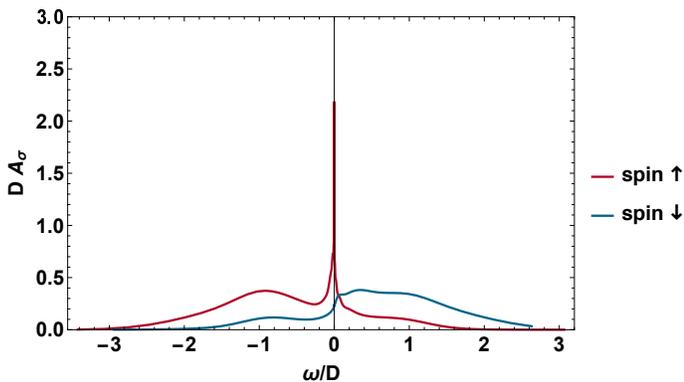}    
\caption{Impurity spectral functions for the excited model at $\delta/\lambda=1/5$. The qualitative form is consistent with expectation, and the Friedel sum rule is satisfied within $\sim20\%$.}
\label{fig:2spec}
\end{figure}

Finally, we discuss  the spectral functions in the valley asymmetric system. A representative plot with both spins together is shown in fig. \ref{fig:2spec}. As stated during the construction of the spectral function, the spin polarization is typically under-estimated unless one employs reduced density matrices, but the qualitative spin polarization is very clear. The spin up chemical potential is higher for $\delta>0$ leading to an effectively deeper impurity level; the resultant increase in occupation appears as a more pronounced bump in $A_{\uparrow}(\omega)$ at negative $\omega$, alongside a very diminished amplitude for positive $\omega$.  The impurity has a higher weight for spin up, but a negligible one for spin down. 

\begin{figure*}[t]
\begin{tabular}{ l | r }
\includegraphics[width=\tabfigwidth]{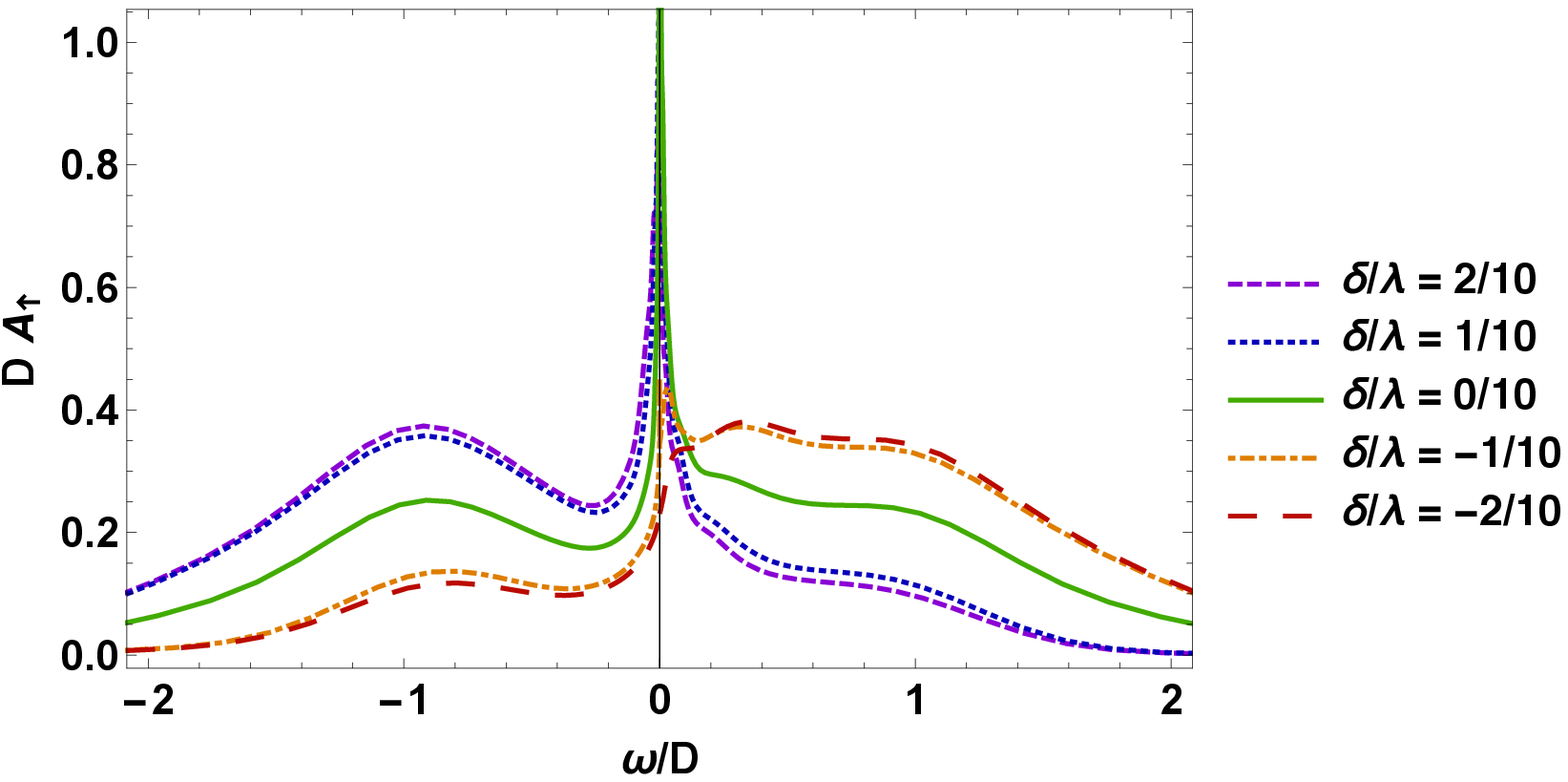}    &
\includegraphics[width=\tabfigwidth]{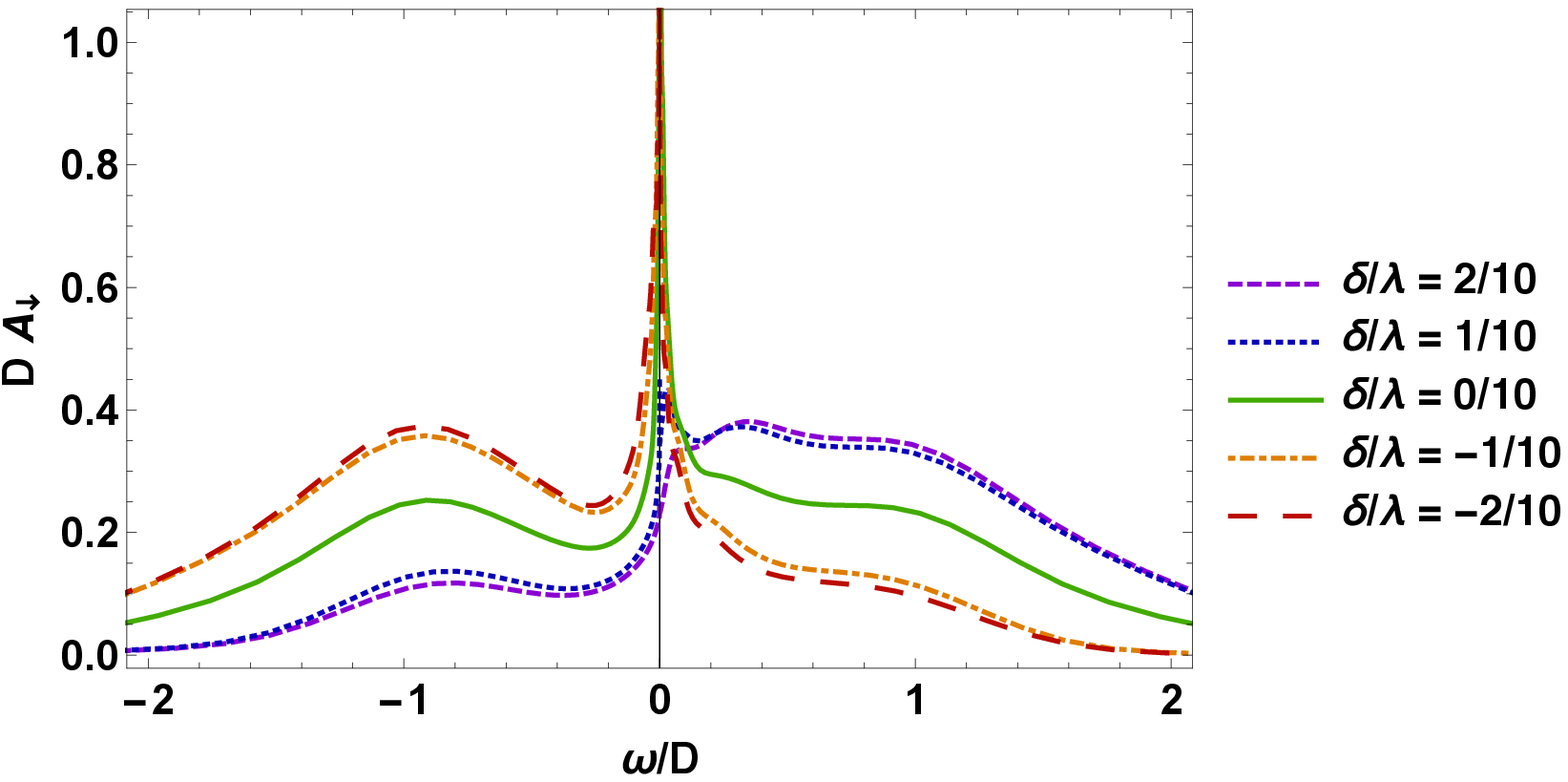} 
\end{tabular}
\caption{The impurity spectral functions at various $\delta$, for spin up (left) and down (right). The pattern of fig. \ref{fig:2spec} is repeated as expected, with a reversal of roles as $\delta$ becomes negative.}
\label{fig:dspec}
\end{figure*}

In fig. \ref{fig:dspec} the spin up/down functions are shown separately, side by side, to illustrate the evolution as a function of the offset $\delta$. The general evolution is consistent with the expectation of larger asymmetry in the spin resolved spectral functions with increasing valley imbalance.


\section{Comparison of Methods}     \label{sec:compare}

The two methods used to explore the Kondo effect in monolayer TMDs, a variational ground state and NRG, provide a detailed understanding of the interplay of topology and interaction in a spin orbit coupled system. The results agree in the case where there is no valley asymmetry. The temperature scale in the variational method (for WSe$_{2}$ with the parameters of the previous section) is $\sim4-8$ K, and in NRG at large $U$ it is $\gtrsim6$ K. The total spin is zero as the solution is an equal admixture of spin singlet and the $m=0$ triplet. While we do not compute $J^{2}$ within NRG, the fact that the susceptibility remains finite at low temperatures is consistent with finite magnetic fluctuations in the ground state, in agreement with the $J^{2}\neq 0$ result of the variational approach. The consistent results provides additional support to the conclusions regarding the unconventional nature of the phenomena\cite{wilsonnrg,kmww1,costinrg,bcpnrg,bullanrg,hewson}.

The lack of a variational ansatz when the chemical potential in the valleys are unequal necessitated an NRG study. Since the wave-function in sec. \ref{sec:varfunc} has net $J_{z}=0$, the finite moment result obtained in sec. \ref{sec:nrg} is not accessible. These results indicate that the $m\neq 0$ triplet states need to be admixed to accurately capture the physics in this case. 

The spectral functions from the NRG method offers more information compared to the variational method. For a system in equilibrium the resonance (see fig. \ref{fig:1spec}) displays rather conventional features for an impurity level in the mixed-valence regime, with its energy level close to the chemical potential. Once the chemical potentials are offset (i.e. $\delta\neq 0$) the spectral functions are split but differ significantly when compared to the effect of a magnetic field\cite{costikmag}. In the latter case the energy of the spin up and spin down states on the impurity site differ with a smaller effect on the Fermi levels, while in the former the spins are degenerate on the impurity site but the Fermi energies are spins split.  The sign of offset acts like the sign of an effective magnetic field as reflected in fig. \ref{fig:dspec} with the appropriate role-reversal of spin up and down. The key difference is the persistence of the sharp peak for one of the spins as $\delta$ is increased. In the presence of a magnetic field the peaks are split and suppressed symmetrically. It is worth noting also that the appropriate sharp peak is shifted slightly in the expected direction as $\delta$ is increased (note the difference in sign between our effective field and the $H$ of ref. \onlinecite{costikmag}).


\section{Conclusions} TMDs provide an exciting new venue to study the interplay of spin orbit coupling, topology and correlations. While prior studies on two dimensional systems have focussed on Rashba spin orbit coupling \cite{malecki,zus} predicting an increased Kondo temperature, a more general analysis in noncentrosymmetric metals showed that the conclusions deduced from them were not universally true \cite{isaev}. In TMDs we have a particular realization of Dresselhaus spin orbit coupling that leads to a lowering of the Kondo scale which arises from the reduced density of states due to spin splitting, reduction of effective band width, and suppression of hybridization due to Berry curvature. The dominant hybridization channel is also determined by the symmetry of the impurity state and the Berry curvature. The finite triplet admixture assures the resonance contains spin fluctuations though it remains nonmagnetic under normal conditions. Furthermore the composition of the Kondo cloud can be tuned by circularly polarized light leading to a (steady-state) optically excited Fermi sea.

To capture the effect of valley asymmetric systems generated by optical excitations, an unbiased NRG approach is implemented. Unlike the effect of a magnetic field which splits the Kondo resonance and suppresses the weight, the time reversal breaking here leads to a preservation of the peak of one of the spin components. From NRG analysis we infer that $m\neq0$ triplet components are admixed when the Fermi sea is given an offset $\delta\neq0$ between the valleys. The  Kondo ground state in TMDs has finite triplet contributions and its magnetic/spin content is tuned by the application of circularly polarized light, opening the door for optomagnetic manipulation and providing a new route for studying Kondo phenomena.

We thank Evan Sosenko, and H.R. Krishna-murthy for useful discussions. VA acknowledges the support of the Hellman foundation,  via the Hellman Faculty Fellowship, and NSF, via grant NSF DMR-1506707.


\begin{appendices}

\section{NRG Procedure}   \label{app:nrg}

The details of the NRG procedure, namely the construction of the linear chain for the appropriate Anderson model, the iterative diagonalization and truncation, are included in this appendix. The zeroth site of the chain represents the state of the conduction electrons that hybridize with the impurity. The chain is constructed in two main steps: first, energy space is discretized into logarithmic intervals approaching the chemical potential, and second the discretized states are superposed to form states orthogonal to the zeroth site. Since the new states are admixtures of the eigenstates of the Hamiltonian without hybridization, they no longer diagonalize the full Hamiltonian. The key insight is that the resulting Hamiltonian is approximated to high accuracy with on site energy and nearest neighbor hopping on the chain (i.e. ``tridiagonalization").  Once Hamiltonians and transformations are set up, the iterative diagonalization procedure is used, allowing for the computation of thermodynamic and spectral properties.

\renewcommand{\thesubsection}{\roman{subsection}}		

\subsection{Logarithmic Discretization}	\label{app:log}

The first step is to recast the QAM basis in  eq. (\ref{eq:type1Hybalt}) in energy space, $\varepsilon$, with the density of states $g(\varepsilon)$, in terms of the dimensionless variable $\xi = \varepsilon / D$.  The Anderson Hamiltonian (\ref{eq:nrgH1}) becomes

\bea
{H' \over D} &=& \sum_{s} \int_{-\ell_{s}}^{e_{s}} d\xi \, \left[ \xi \, c_{\xi,s}\dag c_{\xi,s} + \sqrt{\Gamma \over \pi D} \of{c_{\xi,s}\dag f_{s} + h.c.} \right]  \nonumber  \\
&+& {1 \over D} \sum_{s} \of{\varepsilon_{0,s} + {U \over 2}} n_{f,s} + {U \over 2 D} \of{\sum_{s}n_{f,s}  -  1 }^{2}	\hspace{0.5cm}	
\label{eq:nrgH2}
\eea
with the bottom/top of the rescaled bands $\ell_{s} = 1 + s \delta / 2D$ and $e_{s} = e_{0}/D - s \delta / 2D$, effective impurity levels $\varepsilon_{0,s} = \varepsilon_{0} - s \delta / 2$, and half-width $\Gamma = \pi g(\mu) |v_{\mu}^{\alpha}|^{2}$.

Next, energy space is discretized into logarithmic intervals approaching the chemical potential. This is done to capture the logarithmic divergences that are anticipated at low temperatures --- each interval will contribute an equal amount to integrals like $\int_{-1}^{-k_{B}T/D} d\xi / \xi$.
Intervals are chosen at positive and negative `energy' labeled by $n=0,1,2,...$ approaching zero\cite{kmww1}
\be
e_{s} R^{-(n+1)} < \xi < e_{s} R^{-n}  \; , \;  -\ell_{s} R^{-n} < \xi < -\ell_{s} R^{-(n+1)}
\label{eq:logint}
\ee
where $R>1$ is the discretization parameter\footnotemark[\value{footnote}].
The continuous limit is obtained by taking $R\to1$. The standard choice of $R=3$, known to accurately reproduce the asymptotic results, is made for the results reported here.

Positive and negative energy operators, corresponding to the left/right of (\ref{eq:logint}), are defined as $a_{n,s}$ and $b_{n,s}$:

\bea
a\dag_{n,p,s} &=& \int_{-\ell_{s}}^{e_{s}} d\xi \, \psi_{n,p,s}^{+}(\xi) c\dag_{\xi,s}		\nonumber	\\
b\dag_{n,p,s} &=& \int_{-\ell_{s}}^{e_{s}} d\xi \, \psi_{n,p,s}^{-}(\xi) c\dag_{\xi,s}		\;	.	
\label{eq:abDef}
\eea

The Fourier components $\psi_{n,p,s}^{\pm}(\xi)$ are broken down by intervals, 

\bea
\psi_{n,p,s}^{+}(\xi) &=& \left[{R^{n} \over e_{s} \of{1-R^{-1}}  }\right]^{1/2}  \, \text{e}^{i \omega_{n,s}^{+} p \xi}	\; ,	\nonumber	\\ 
&& \text{for}	 \;\,	e_{s} R^{-(n+1)} < \xi < e_{s} R^{-n}	\; ; 	\nonumber	\\
\psi_{n,p,s}^{-}(\xi) &=& \left[{R^{n} \over \ell_{s} \of{1-R^{-1}}  }\right]^{1/2}  \, \text{e}^{-i \omega_{n,s}^{-} p \xi}	\; ,	\label{eq:logFourier}	\\
&& \text{for}	 \;	 -\ell_{s} R^{-n} < \xi < -\ell_{s} R^{-(n+1)} \nonumber
\eea

with fundamental frequencies $\omega_{n,s}^{+} = 2\pi R^{n}/e_{s}(1-R^{-1})$, $\omega_{n,s}^{-} = 2\pi R^{n}/\ell_{s}(1-R^{-1})$, and integral harmonic index $p\in(-\infty,\infty)$. The components $\psi^{\pm}$ are zero when the energy $\xi$ lies outside the respective $n^{th}$ intervals on the r.h.s. above.  In terms of this complete set of properly orthonormalized states one obtains

\be
c\dag_{\xi,s} = \sum_{n,p} \left[  \of{\psi_{n,p,s}^{+}(\xi)}^{\star} a\dag_{n,p,s}   +   \of{\psi_{n,p,s}^{-}(\xi)}^{\star} b\dag_{n,p,s}    \right]
\label{eq:cabDef}
\ee

leading to a chain Hamiltonian, for the TMD electrons only,

\begin{widetext}
\bea
\label{eq:fourierH}
&& \int_{-\ell_{s}}^{e_{s}} d\xi \,\xi c\dag_{\xi,s} c_{\xi,s} = \\
 &&{1 + R^{-1} \over 2} \sum_{n,p} \of{ e_{s} R^{-n} a_{n,p,s}\dag a_{n,p,s}   - \ell_{s} R^{-n} b_{n,p,s}\dag b_{n,p,s} }  
+	{1 - R^{-1}  \over  2\pi i}  \sum_{n,p' \neq p} \left[	{R^{-n} \over p' - p} \text{e}^{2\pi i (p' - p)} \of{ e_{s} a_{n,p,s}\dag a_{n,p',s} - \ell_{s} b_{n,p,s}\dag b_{n,p',s} }	\right]	.	\nonumber
\eea
\end{widetext}

The band Hamiltonian is diagonal in the original basis $c\dag_{\xi,s}$. The use of any other basis will inevitably lead to off-diagonal terms like those in the final sum above. Notice, however, that the off-diagonal sum comes with a factor $\propto 1-R^{-1}$ so it vanishes as the discretization factor $R$ approaches unity (i.e. as continuity is restored). With $R=3$, the suppression factor is $\propto 2/3$. Following the standard procedure \cite{kmww1} the \emph{crucial approximation} of using only harmonic index $p=0$ is implemented eliminating the off-diagonal sum. The components (\ref{eq:logFourier}) are now constants (of varying magnitude) in each energy interval; in other words the states created by $a\dag_{n,0,s}$ and $b\dag_{n,0,s}$ are simple (normalized) averages of the states $c\dag_{\xi,s}$ on the appropriate intervals of energy $\xi$. Thus the approximation of retaining only these states is equivalent to disregarding fluctuations across individual energy intervals, only allowing changes across interval boundaries. Since phase changes like the Kondo effect are known to exhibit changes across the entire energy space (and physical space), Wilson's approximation of taking only the constant $p=0$ contribution is justified. The Hamiltonian takes the form, 
\bea
{H' \over D} &\approx& {1 + R^{-1} \over 2} \sum_{n,s} \of{ e_{s} R^{-n} a_{n,s}\dag a_{n,s}   - \ell_{s} R^{-n} b_{n,s}\dag b_{n,s} }	\nonumber	\\
&& + {{H}_{V}  \over D} + {H_{imp}  \over D}	.
\label{eq:nrgH3}
\eea

\subsection{Tridiagonalization}  	\label{app:tridiag}

The next step is to connect eq. (\ref{eq:nrgH3}) above to eq. (\ref{eq:nrgH4}) of sec. \ref{sec:nrgsetup}. 
In the previous section the band (TMD) operators have been appropriately discretized to focus on states near the chemical potentials. While this is (approximately) diagonal, the coupling to impurity involves a different state:
\bea
\int_{-\ell_{s}}^{e_{s}} d\xi \, c_{\xi,s}\dag &=& \sum_{n} \sqrt{1-R^{-1} \over R^{n}} \left[ (e_{s})^{1/2} a_{n,s}\dag  +  (\ell_{s})^{1/2} b_{n,s}\dag \right] 	.
\nonumber \\
&\equiv& \of{1+e_{0}/D}^{1/2} d_{0,s}\dag		.
\label{eq:site0}
\eea
The state created by $d_{0,s}\dag$ is interpreted as the zeroth site of a semi-infinite chain, which has the impurity coupled to it. 

The task is now to create other states which are orthogonal to the zeroth site (and to each other) in such a way that nothing further than nearest-neighbor hoppings appear in the final chain Hamiltonian (see the first line of eq. (\ref{eq:nrgH4})). In this section  $H_{c}$  denotes the $p=0$ band Hamiltonian, the first sum of eq. (\ref{eq:fourierH}) above or the first line of eq. (\ref{eq:nrgH3}), which will become the chain Hamiltonian. 

Constructing the chain states $d\dag_{m,s}$ with $m>0$ involves a Graham-Schmidt orthogonalization procedure starting from the zeroth site, implemented using the Lanczos algorithm\cite{hewson,costinrg,bcpnrg}. Dropping the spin label for now, the task is to find a transformation, a set of $\set{u_{m,n}, v_{m,n}}$ with $m,n=0,1,2,...$ describing the relation $d\dag_{m} = \sum_{n}\of{u_{m,n}a\dag_{n} + v_{m,n}b\dag_{n}}$, such that the resultant states are assigned some energies $\epsilon_{m}$ and hoppings $t_{m}$ leading to the tridiagonal form (\ref{eq:nrgH4}). The starting point (using spin labels where appropriate) for $m=0$ is

\be
u_{0,n} = \of{e_{s} (1-R^{-1}) \over 1 + e_{0}/D }^{1/2} R^{-n/2}	\; , \;	v_{0,n} = \of{\ell_{s} \over e_{s}}^{1/2} u_{0,n}
\label{eq:cpars0}
\ee

from the definition of the zeroth site, eq. (\ref{eq:site0}). Consider the zeroth site with state denoted  by $\state{0} = d\dag_{0}\state{vac.}$. The energy of this state is  $\epsilon_{0} = \cstate{0} H_{c} \state{0}$. The next site is given by

\be
\state{1} = { 1 \over t_{0} } \of{ H_{c} \state{0} - \state{0}\cstate{0} H_{c} \state{0} }
\label{eq:csite1}
\ee

which is automatically orthogonal to $\state{0}$ and whose normalization constant $t_{0}$ is the same as the hopping $\cstate{1} H_{c} \state{0}$. In this case the hopping (squared) is also the same as the variance in energy, $t_{0}^{2} = \cstate{0} H_{c}^{2} \state{0} - \cstate{0} H_{c} \state{0}^{2}$. The energy of the new site is $\epsilon_{1} = \cstate{1} H_{c} \state{1} = (\cstate{0} H_{c}^{3} \state{0} - 2 \epsilon_{0} \cstate{0} H_{c}^{2} \state{0} + \epsilon_{0}^{3}  )/t_{0}^{2}$. It is evident that, in constructing many sites, arbitrary powers $\cstate{0} H_{c}^{P} \state{0}$ are needed but every quantity is given in terms of the starting point values (\ref{eq:cpars0}).
Continuing, the next site $m=2$ is given by 

\bea
\state{2} &=& { 1 \over t_{1} } \of{ H_{c} \state{1} - \state{1}\cstate{1} H_{c} \state{1} - \state{0}\cstate{0} H_{c} \state{1} }	\nonumber	\\
&=&	{ 1 \over t_{1} } \of{ H_{c} \state{1} - \state{1}\epsilon_{1} - \state{0}t_{0} }
\label{eq:csite2}
\eea

which is orthogonal to both previous sites $m=0, 1$. Importantly the above state does not connect to the zeroth site: $\cstate{2} H_{c} \state{0} = 0$. This is the key point that keeps the Hamiltonian $H_{c}$ tridiagonal instead of having arbitrarily off-diagonal pieces. The expressions for $\epsilon_{2}$ and $t_{1}$ are messier and involve larger powers of $H_{c}$ in the zeroth state. For a general site $m$,

\be
\state{m+1}  = { 1 \over t_{m} } \of{ H_{c} \state{m} - \state{m} \epsilon_{m} - \state{m-1} t_{m-1} }	.
\label{eq:csitem}
\ee

Using the definitions $\epsilon_{m} = \cstate{m} H_{c} \state{m}$ and $t_{m-1} = \cstate{m-1} H_{c} \state{m}$, the recursion equations that determine the tridiagonal transformation are

\bea
\label{eq:calgo}
t_{m}^{2} &=& \of{ 1 + R^{-1} \over 2}^{2} \sum_{n=0}^{\infty} \left[ |u_{m,n}|^{2} (e_{s} R^{-n})^{2} 	\right.	 \\
&&+ \left. |v_{m,n}|^{2} (\ell_{s} R^{-n})^{2}  \right]  - \epsilon_{m}^{2} - t_{m-1}^{2}		\nonumber	\\
\epsilon_{m} &=& { 1 + R^{-1} \over 2} \sum_{n=0}^{\infty} \left[ |u_{m,n}|^{2} (e_{s} R^{-n})  -  |v_{m,n}|^{2} (\ell_{s} R^{-n}) \right]	\nonumber	\\
u_{m+1,n} &=&  {1 \over t_{m} } \left[ \of{ { 1 + R^{-1} \over 2} e_{s} R^{-n} - \epsilon_{m} } u_{m,n} \right.	\nonumber	\\
&& - \left. t_{m-1} u_{m-1,n}   \right]   	\nonumber	\\
v_{m+1,n} &=&  {1 \over t_{m} } \left[ - \of{ { 1 + R^{-1} \over 2} \ell_{s} R^{-n} + \epsilon_{m} } v_{m,n}  \right.	\nonumber	\\
&& - \left. t_{m-1} v_{m-1,n}   \right]	\; .	\nonumber
\eea

The above equations are consistent with those of section II C of ref. \onlinecite{bcpnrg}. They are valid for all sites, $m\geq0$, as long as one takes $t_{-1}\to0$. For the symmtric Anderson model, the site energies $\epsilon_{m}$ are zero and Wilson has constructed\cite{wilsonnrg} an analytic solution for the hoppings $t_{m}$. When the chemical potential is placed asymmetrically in a band the site energies are no longer zero, and the energies and hoppings must generally be determined numerically from (\ref{eq:calgo}). For numerical expedience it is useful to exploit the equivalent relations for powers, i.e. for $\cstate{m} H_{c}^{P} \state{m+1}$ and $\cstate{m+1}H_{c}^{P} \state{m+1}$ rather than perform infinite sums as in (\ref{eq:calgo}). The energies and hoppings are then obtained with $P=1$. Either way, one can numerically calculate the tridiagonal chain parameters up to any arbitrary chain length by iterating a set of equations like (\ref{eq:calgo}). The procedure must be repeated for both spin up and down, at least when any spin dependence is involved. If there is more than one channel coupled to the impurity, that is if one effectively has multiple chains (more labels than spin), then the procedure is to be repeated for each of those as well.

\subsection{Iterative Diagonalization}	\label{app:iterdiag}

Starting from the semi-infinite chain Hamiltonian (\ref{eq:nrgH4}), in this section the rescaled Hamiltonian for finite systems, $\overline{H}_{M}$ from eq. (\ref{eq:nrgHlim}) is constructed and diagonalized. A finite chain up to site $M$ is considered. The nearest neighbor hoppings scale as $R^{-m/2}$, so the Hamiltonian is rescaled by a factor $\propto R^{M/2}$ to ensure each new site enters the system with energy of order unity\cite{wilsonnrg,kmww1}. The Hamiltonian is

\begin{widetext}
\bea
\overline{H}_{M} &=& { 2 \over 1 + R^{-1} } R^{(M-1)/2}  \left[ \sum_{s} \of{ \sum_{m=0}^{M} \epsilon_{m,s} d_{m,s}\dag d_{m,s} + \sum_{m=0}^{M-1} t_{m,s} \of{d_{m+1,s}\dag d_{m,s} + h.c.} }   \right. \nonumber \\
&& + \left. \widetilde{\Gamma}^{1/2} \sum_{s} \of{d_{0,s}\dag f_{s} + h.c.}  +  \sum_{s} \widetilde{\delta}_{f,s} n_{f,s} + \widetilde{U} \of{\sum_{s} n_{f,s} -1}^{2}   \right]
\label{eq:nrgH5}
\eea
\end{widetext}
where $\widetilde{\Gamma} = \Gamma (1+e_{0}/D) / \pi D$, $\widetilde{\delta}_{f,s} = (\varepsilon_{0,s} + U/2)/D$, and $\widetilde{U}=U/2D$.  
By construction, the full semi-infinite Hamiltonian is recovered upon taking the limit as in eq. (\ref{eq:nrgHlim}).
The recursion relation needed to build the $M+1$ Hamiltonian is

\bea
\overline{H}_{M+1} &=& R^{1/2} \overline{H}_{M}  +  { 2 \, R^{M/2} \over 1 + R^{-1} }  \sum_{s} \left(  \epsilon_{M+1,s} d_{M+1,s}\dag d_{M+1,s} \right.	\nonumber	\\
&& + \left. t_{M,s} \of{d_{M+1,s}\dag d_{M,s} + h.c.}  \right)	.
\label{eq:nrgHrecur}
\eea

The general methodology of the iterative procedure is as follows. For details see  refs. \onlinecite{bcpnrg,kmww1}. Begining with the simplest chain consisting only of the impurity and the zeroth chain site coupled to it ($\overline{H}_{0}$), the system is numerically diagonalized in Fock space, broken down by subspaces. To label the subspaces, note that the number of spin up/down is conserved; the operators $N_{s} = n_{f,s} + \sum_{m} d_{m,s}\dag d_{m,s}$ commute with the Hamiltonian. For convenience, the subspace labels of `charge' $Q = \sum_{s}N_{s} - (M+2)$ and spin $J_{z} = (1/2) (N_{\uparrow} - N_{\downarrow})$ are chosen.

Once the initial $M=0$ Hamiltonian is diagonalized in each subspace, the $M=1$ Hamiltonian is constructed by extending the previous eigenspace to include the new site and using eq. (\ref{eq:nrgHrecur}). The main drawback with this approach is that the Hilbert space grows by a factor of 4 with each additional site; the total number of states to describe up to site $M$ is $4^{M+2}$, since each site has spin up/down which can be either empty or filled. To deal with this, a choice of a number of states to keep  is made and all states above the energy of the last state (keeping degenerate states) are discarded. The number of states to be kept is related to the choice of discretization parameter $R>1$; approaching unity is equivalent to taking the  continuity limit implying that a large number of states is to be included to describe the system accurately. Typically, a choice of $R=3$ requires about 400 states whereas $R=2.5$ requires 600 or more. The rate of convergence is also influenced by $R$ (larger is faster). For the sake of time and memory usage a choice of $R=3$ and 400-425 states is used in this work.

With the truncation step the procedure of diagonalization, truncation, and Hamiltonian construction  is implemented until a sufficiently large site $M$ is reached. The number of steps is related to convergence which in turn is controlled by $R$. To get accurate results the iteration is stopped once the system has stabilized, as revealed by vanishing differences in the energy spectrum in going from $M$ to $M+2$. The finite-size effects create oscillations for even/odd total sites and for thermodynamic results (next) an interpolation and averaging of the even/odd results are implemented.

\subsection{Thermodynamics and Spectral Function}		\label{app:results}

Here we describe the procedure to compute thermodynamics properties and the spectral function.
Since the states are truncated in the NRG scheme, each step introduces an error in calculated states, and higher energies are neglected by choosing a lower temperature. Denoting the maximum energy scale of the finite system $K(R)$, the replacement of $Z$ by $Z_{M}$ is a good approximation provided that \cite{costinrg}

\be
1/K(R)  \ll  \overline{\beta}_{M}  \ll R	.
\label{eq:nrgbeta}
\ee

For $R=3$, $K(R) \sim 10$, thus fixing $\overline{\beta}_{M} = 1/2$ satisfies the necessary condition. In this case, the effective temperatures considered are decreasing exponentially with the number of sites $M$ included: $k_{B}T_{M}/D = (1 + R^{-1}) R^{-(M-1)/2}$.

Thermodynamic averages are found in the usual way, as in eq. (\ref{eq:impocc}), but matrix elements are determined based on the symmetries of the system. Since the system is broken down to `charge' $Q$ and total spin $J_{z}$ subspaces, their averages are readily evaluated --- sums such as (\ref{eq:impocc}) are calculated in each subspace (each with fixed $Q$, $J_{z}$), which are then totaled. The same can be done for powers, e.g. $J_{z}^{2}$.

To analyze the nature of the low-temperature system, i.e. the formation of a bound singlet state, the impurity contributions to entropy and susceptibility are needed. As remarked in sec. \ref{sec:thermo}, to calculate impurity contributions the results for the system as a whole are first computed and then the part arising from the band (TMD) alone without the impurity is subtracted\cite{bcpnrg,wilsonnrg}:

\bea
S_{imp} = S_{total} - S_{b}	\nonumber	\\
\chi_{imp} = \chi_{total} - \chi_{b}	.
\label{eq:impentsus}
\eea
To compare the spin $J_{z}$ to the variational results, the spin of the chain without the impurity $\avg{J_{z}} - \avg{J_{z}}_{b}$ is subtracted, but the same quantity is accurately given by the impurity spin $\avg{ s_{z}}$ as expected.

To avoid taking derivatives of the free energy (to minimize numerical errors) the alternative form of entropy is employed 

\be
S_{total}/k_{B} = \beta \langle H' \rangle + \ln Z = \overline{\beta}_{M} \langle \overline{H}_{M} \rangle + \ln Z_{M},	
\label{eq:entdef}
\ee
where the average energy is  $\langle \overline{H}_{M} \rangle = \sum_{q} \overline{E}_{q} \exp\of{-\overline{\beta}_{M} \overline{E}_{q} } / Z_{M}$.  For the susceptibility, the fluctuation-dissipation theorem is used to write
\be
\chi_{total}/(g\mu_{B})^{2} = \beta \of{ \langle J_{z}^{2} \rangle  -  \langle J_{z} \rangle^{2}	}	.
\label{eq:susdef}
\ee
As stated above, the impurity contributions are found by subtracting the values arising from the system with no impurity site. While chain only values can by  computed by hand, for numerical consistency a separate NRG procedure for the chain without the impurity is utilized. 

The complication that remains is the even/odd oscillations arising from finite-size effects. The scheme used here is to find separate interpolation functions for the two cases and average them. The procedure is accurate at low temperatures but boundary effects render the highest temperature point(s) unreliable. 


In addition to the thermodynamic properties, the spectral functions, defined in eq. (\ref{eq:spectdef}), provide information on the spin content of the resonance.
Focusing on the ground-state properties the $T=0$ spectral function is computed. Only states differing from the ground state by one particle and one unit of spin, i.e. $|Q_{q} - Q_{0}| = 1$ and $|J_{z,\,q} - J_{z,\,0}| = 1/2$, need to be considered. To compute the spectral function for $T\neq0$ combinations between \emph{any} states, not just the ground state, with such differences in $Q$ and $J_{z}$ need to be considered making the following construction more involved\cite{bullanrg}. 

In order to construct a spectral function within NRG the matrix elements of impurity operators $f_{s}$ projected to the eigenspace of each successive chain are required. Additionally, the energy range must be restricted for each $M$ since each iteration reduces the energy window in which results are accurate. The restriction cuts off the low energies which are yet to be obtained from larger $M$ and cuts out the high energies which are already accurately described from previous $M$. Setting $\omega_{M}=((1 + R^{-1}) / 2) R^{-(M-1)/2} = \overline{\beta}_{M}/(D \beta)$, the range is $\omega_{M} < |\omega| < K(R) \omega_{M}$.

The energy ranges for different $M$ overlap and the spectral function amplitudes must be accurately calculated without double counting while maintaining the accuracy of low-energy states. There is a well-defined method for this purpose \cite{bullanrg}, which is described as follows. Consider chains $M$ and $M+2$ (recall that even and odd are separated due to finite size effects): the overlap region will involve the higher energies from $M+2$ and the lower energies from $M$. Since its accuracy is at lower energy, the $M+2$ spectral function is weighted in the overlap region with a linear distribution going from one down to zero towards higher energy. Conversely, the $M$ spectral function is also weighted in the overlap region but it is zero at the lower end and linearly increases to one at higher energy. After adding the weighted $M$ and $M+2$ spectral functions, accurate results for  a larger energy range with appropriate emphasis and without double counting (the sum of the linear distributions is unity) are obtained. Thus the  total spectral function covering the full energy range is built by adding all even/odd functions up to a maximum chain length. To complete the total function the even and odd results are averaged, $A=(A^{even}+A^{odd})/2$.

 The $\delta$-peaks appearing in (\ref{eq:spectdef}) are implemented by using a broadening function,

\be
\delta\of{x} \to {1 \over \sqrt{\pi \eta_{M}^{2}} } \exp\of{-x^{2}/\eta_{M}^{2} }
\label{eq:spectspread}
\ee
where the widths are $\eta_{M} \approx \omega_{M}$.
This allows for the determination of smooth spectral functions for the impurity.  When plotting these, because of the exponentially decreasing widths, it is beneficial to sample the energies logarithmically so that most sample points lie near $\omega=0$.

\end{appendices}


\end{document}